\documentclass[12pt]{article}
\usepackage{epsfig}
\usepackage{afterpage}
\usepackage{float}

\def\etal{{\it et al.}}

\begin{document}
%
\vspace{1.5in}
\title{$\,$~~~~~~~~~~~~~~~~~~~~~~~~~~~~~~~~~~~~~~~~~~~~~~~~~~~~
${B}\to D^*\pi^+\pi^-\pi^-\pi^o$, $D^{(*)}\omega\pi^-$ and the 
Observation of a Wide $1^-$ $\omega\pi^-$ Enhancement at 1418 MeV} 

\author{ CLEO Collaboration}

\date{June 15, 2000}
\vspace{1.in}
\maketitle
\vspace{-2.7in}
\vbox{\rightline{\hfil CLEO CONF 00-1}
\rightline{\hfil ICHEP00-78~~~~~~~~}} 
\vspace{2.5in}
\vspace{1in}
\begin{abstract} 
We report on the observation of  ${B}\to D^{*}\pi^+\pi^-\pi^-
\pi^o$ decays. 
The branching ratios for $D^{*+}$ and $D^{*o}$ are  
(1.72$\pm$0.14$\pm$0.24)\%
and (1.80$\pm$0.24$\pm$0.25)\%,
respectively. Each final state has a 
$D^{*}\omega\pi^-$ component, with branching ratios 
(0.29$\pm$0.03$\pm$0.04)\% and (0.45$\pm$0.10$\pm$0.07)\% for the 
$D^{*+}$ and $D^{*o}$ modes, respectively.  We also observe 
${B}\to D\omega\pi^-$ decays.   
The branching ratios for $D^{+}$ and $D^{o}$ are  
(0.28$\pm$0.05$\pm$0.03)\% and (0.41$\pm$0.07$\pm$0.04)\%,  
respectively. The $\omega\pi^-$ appears to come from the decay of 
a wide $1^-$ resonance. A fit to a Breit-Wigner shape gives a 
mass of 1418$\pm$26$\pm$19 MeV and width of 388$\pm$44$\pm$32 MeV.
We identify this object as the $\rho$(1450) or $\rho'$.   
\end{abstract}
\vfill
\begin{flushleft}
.\dotfill .
\end{flushleft}
\begin{center}
Submitted to XXXth International Conference on High Energy Physics, July 2000,
Osaka, Japan
\end{center}

\newpage
\begin{center}
M.~Artuso,$^{1}$ R.~Ayad,$^{1}$ C.~Boulahouache,$^{1}$
K.~Bukin,$^{1}$ E.~Dambasuren,$^{1}$ S.~Karamov,$^{1}$
G.~Majumder,$^{1}$ G.~C.~Moneti,$^{1}$ R.~Mountain,$^{1}$
S.~Schuh,$^{1}$
% ...
\end{center}
 
\small
\begin{center}
$^{1}${Syracuse University, Syracuse, New York 13244}\\
% ...
\end{center}

\newpage
{\renewcommand{\thefootnote}{\fnsymbol{footnote}}
\setcounter{footnote}{0}
}

\section{Introduction}\label{sec:Introduction}

Understanding hadronic decays of the $B$ is crucial to insuring 
that decay modes used for measurement of CP violation truly 
reflect the underlying quark decay mechanisms expected 
theoretically.
 
Currently, measured exclusive branching ratios for hadronic $B$ 
decays total only a small fraction of the hadronic width. The 
semileptonic branching
ratio for $B\to Xe^-\nu$, $X\mu^-\nu$, and $X\tau^-\nu$ totals 
approximately
25\% \cite{PDG}. The measured hadronic decay modes for the 
$\overline{B}^o$
including $D^+ (n\pi^-)$, $D^{*+} (n\pi^-)$, where $3\ge n \ge 1$, 
$D^{+(*)}D_s^{-(*)}$, and $J/\psi$ exclusive totals only about 
10\% \cite{PDG}. (The $B^-$
modes total about 12\%.) Thus our understanding of hadronic $B$ 
decay modes is not yet well based in data.

It is also interesting to note that
the average charged multiplicity in a hadronic $B^o$ decay is 
5.3$\pm 0.1$ 
\cite{chargedmult}. Since this multiplicity contains 
contributions from the
$D^+$ or $D^{*+}$ normally present in $\overline{B}^o$ decay, we 
expect a sizeable, approximately several percent, 
decay rate into final states with four pions \cite{Argus}.
The seen $D^{(*)} (n\pi)^-$ final states for $n \leq 3$ are 
consistent with 
being quasi-two-body final states. For $n$ of two the $\rho^-$ 
dominates, while
for $n$ of three the $a_1^-$ dominates \cite{BigB}. These decays 
appear to occur from a simple spectator mechanism where the 
virtual $W^-$ materializes as a single hadron: $\pi^-$, $\rho^-$ 
or $a_1^-$. 

In this paper we investigate final states for $n$ of 
4. We will show a large signal for the $D^{*+}\pi^+\pi^-\pi^-\pi^o$ 
final state in section~\ref{sec:Dstarp4pi}. In section~\ref{sec:Dspomegapi}
we will show that a substantial 
fraction, $\sim$20\%
arise from $D^{*+}\omega\pi^-$ decays and that the $\omega\pi^-$ mass
distribution has a resonant structure around 1.42 GeV with a width
of about 0.4 GeV. In section~\ref{sec:Ds04pi}, the similar conclusions are
drawn about the $D^{*o}\pi^+\pi^-\pi^-\pi^o$ final state.
 The same structure is shown to exist  in
$D\omega\pi^-$ final states (section~\ref{sec:Domegapi}) and we will use these
events to show in section~\ref{sec:Dangular} that
the spin-parity is most likely $1^-$. This state is identified as the 
$\rho'$, also sometimes called the $\rho$(1450).
Other resonant substructure is searched for, but not found
(section~\ref{sec:nullsearch}). Finally we summarize our findings and compare
with the predictions of factorization and other models in section~\ref{sec:conclusions}. 

The data sample consists of 9.0 fb$^{-1}$ 
of integrated luminosity taken with the CLEO II and II.V
detectors \cite{CLEOdetector} using the CESR $e^+e^-$ storage ring 
on the peak of the 
$\Upsilon(4S)$ resonance and 4.4 fb$^{-1}$ in the
continuum at 60 MeV less center-of-mass energy. The sample 
contains
19.4 million $B$ mesons.

\section{Common Selection Criteria}\label{sec:Common}

Hadronic events are selected by a minimum of five charged tracks, total
visible energy greater than 15\% of the center-of-mass energy, and a charged
track vertex consistent with the nominal interaction point.
To reject continuum we require that the Fox-Wolfram moment $R_2$ 
be less than 0.3 \cite{Fox-Wolf}. 

Track candidates are required to pass through a common spatial 
point defined by origin of all tracks. Tracks with momentum below 
900 MeV/c are required to have ionization loss in the drift 
chamber within $3\sigma$ of their assigned mass.\footnote{Here and throughout
this paper $\sigma$ indicates an r.m.s. error.} (These 
requirements are not imposed on slow charged pions from $D^{*+}$ 
decay.) Photon candidates are required
to be within the ``good barrel region," within 45$^{\circ}$ of 
the normal to the beam line, and have an energy distribution in 
the CsI calorimeter consistent with that of an electromagnetic 
shower. To select $\pi^o$'s,
we require that the diphoton invariant mass be between  -3.0 to 
+2.5$\sigma$, where $\sigma$ varies with momentum and has an
average value of approximately 5.5 MeV. For each two-photon mass combination
$\sigma$ is calculated. After candidate selection the two-photon's are
kinematically fit by constraining their invariant mass to that of the $\pi^o$.

We select $D^o$ and $D^+$ candidates via the decay modes shown in 
Table~\ref{table:massres}. 
We require that the invariant mass of the $D$  candidates lie 
within $\pm 2.5\sigma$ of the known $D$ masses. The $\sigma$'s are 
also
listed in Table~\ref{table:massres}. The $D^o$ widths vary with 
the  
$D^o$ momentum, $p$, (units of MeV).   

We select $D^{*+}$ candidates by imposing the addition requirement
that the mass difference between $\pi^+ D^o$ and 
$D^o$ combinations is within $\pm 2.5\sigma$ of the known mass 
difference.
For the $D^{*o}$, we use the same requirement for the $\pi^o D^o$ decay. The mass 
difference resolutions are 
0.63 and 0.90 MeV, for the $\pi^+D^o$ and $\pi^o D^o$ modes, 
respectively\cite{Differences}

\begin{table}[hbt]
\begin{center}
\caption{Mass Resolutions ($\sigma$) in MeV}
\label{table:massres}
\begin{tabular}{cccc}\hline\hline
$D^+\to K^-\pi^+\pi^+$ &  $D^o\to K^-\pi^+$ & 
$D^o\to K^-\pi^+\pi^o$ & $D^o\to K^-\pi^+\pi^+\pi^-$ \\
6.0 & $p\times$0.93$\times 10^{-3}$+6.0 & 
$p\times$0.68$\times 10^{-3}$+11.6 & 
$p\times$0.92$\times 10^{-3}$+4.7 \\\hline
\end{tabular}
\end{center}
\end{table}

\section{Observation of $\overline{B}^o\to D^{*+}\pi^+\pi^-\pi^-
\pi^o$ Decays}\label{sec:Dstarp4pi}

\subsection {$B$ Candidate Selection}

We start by investigating the $D^{*+}(4\pi)^-$ final 
state.\footnote{In this paper $(4\pi)^-$ will always denote the 
specific combination $\pi^+\pi^-\pi^-\pi^o$.} 
The $D^{*+}$  candidates are combined with all combinations of
$\pi^+\pi^-\pi^-\pi^o$ mesons. 

Next, we
calculate the difference between the beam energy, $E_{beam}$, and 
the measured energy
of the five particles, $\Delta E$. The ``beam constrained" 
invariant mass of the $B$
candidates, $M_B$, is computed from the formula
\begin{equation}
M_B^2=E_{beam}^2-(\sum_i\overrightarrow{p_i})^2~~~.
\end{equation}

To further reduce backgrounds we define
\begin{equation}\label{eq:chisq}
\chi_b^2=\left({{\Delta M_{D^*}}\over {\sigma(\Delta 
M_{D^*})}}\right)^2 + 
\left({{\Delta M_{D}}\over {\sigma(\Delta M_{D})}}\right)^2 + 
\sum_{n(\pi^o)}\left({{\Delta M_{\pi^o}}\over {\sigma(\Delta 
M_{\pi^o})}}\right)^2~~~,
\end{equation}
where $\Delta M_{D^*}$ is the computed $D^*-D^o$ mass difference
minus the nominal value, $\Delta M_{D}$ is
the invariant candidate $D^o$ mass minus the known $D^o$ mass and 
$\Delta M_{\pi^o}$ is the measured $\gamma\gamma$ invariant mass 
minus the
known $\pi^o$ mass. All $\pi^o$'s in the final state are included 
in the sum.
The $\sigma$'s are the measurement errors. We select candidate 
events in each mode 
requiring that  $\chi^2_b < C_n$, where $C_n$
varies for each decay $D^o$ decay mode. For the $Kn\pi$ decay 
modes we use
$C_n=~12,$ 8, and 6, respectively.

\begin{figure}[bht]
\centerline{\epsfig{figure=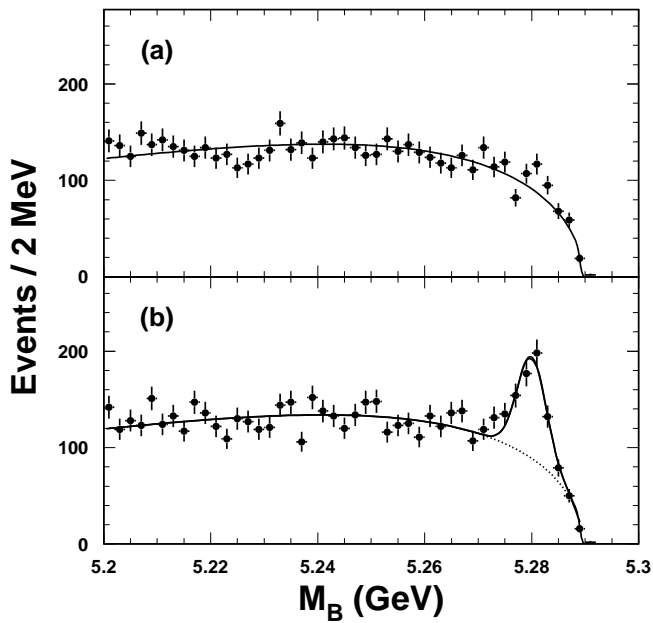,height=4in}}
\vspace{-1.0cm}
\caption{\label{bm_4pi_kpi}The $B$ candidate mass spectra for the
final state
$D^{*+}\pi^+\pi^-\pi^-\pi^o$, with $D^o\to K^-\pi^+$ (a) for 
$\Delta E$
sidebands and (b) for $\Delta E$ consistent with zero. The curve 
in (a) is a
fit to the background distribution described in the text, while in 
(b) the
shape from (a) is used with the normalization allowed to float and 
a signal
Gaussian of width 2.7 MeV is added.}
\end{figure}

\subsection{Branching Fraction and $(4\pi)^-$ Mass Spectrum}

We start with the $D^o\to K^-\pi^+$ decay mode.
We show the candidate $B$ mass distribution, $M_B$, for 
$\Delta E$ in the side-bands from -5.0 
to -3.0$\sigma$
and 5.0 to 3.0$\sigma$
 on Fig.~\ref{bm_4pi_kpi}(a). The $\Delta E$ resolution
is 18 MeV ($\sigma).$ This gives a good representation of the 
background in the signal 
 region.
We fit this distribution with a shape given as 
\begin{equation}
back(r)=p_1 r\sqrt{1-r^2}e^{-p_2(1-r^2)}~~~,
\end{equation}
where $r=M_B/5.2895$, and the $p_i$ are parameters given by the 
fit.

\begin{figure}[hbt]
\centerline{\epsfig{figure=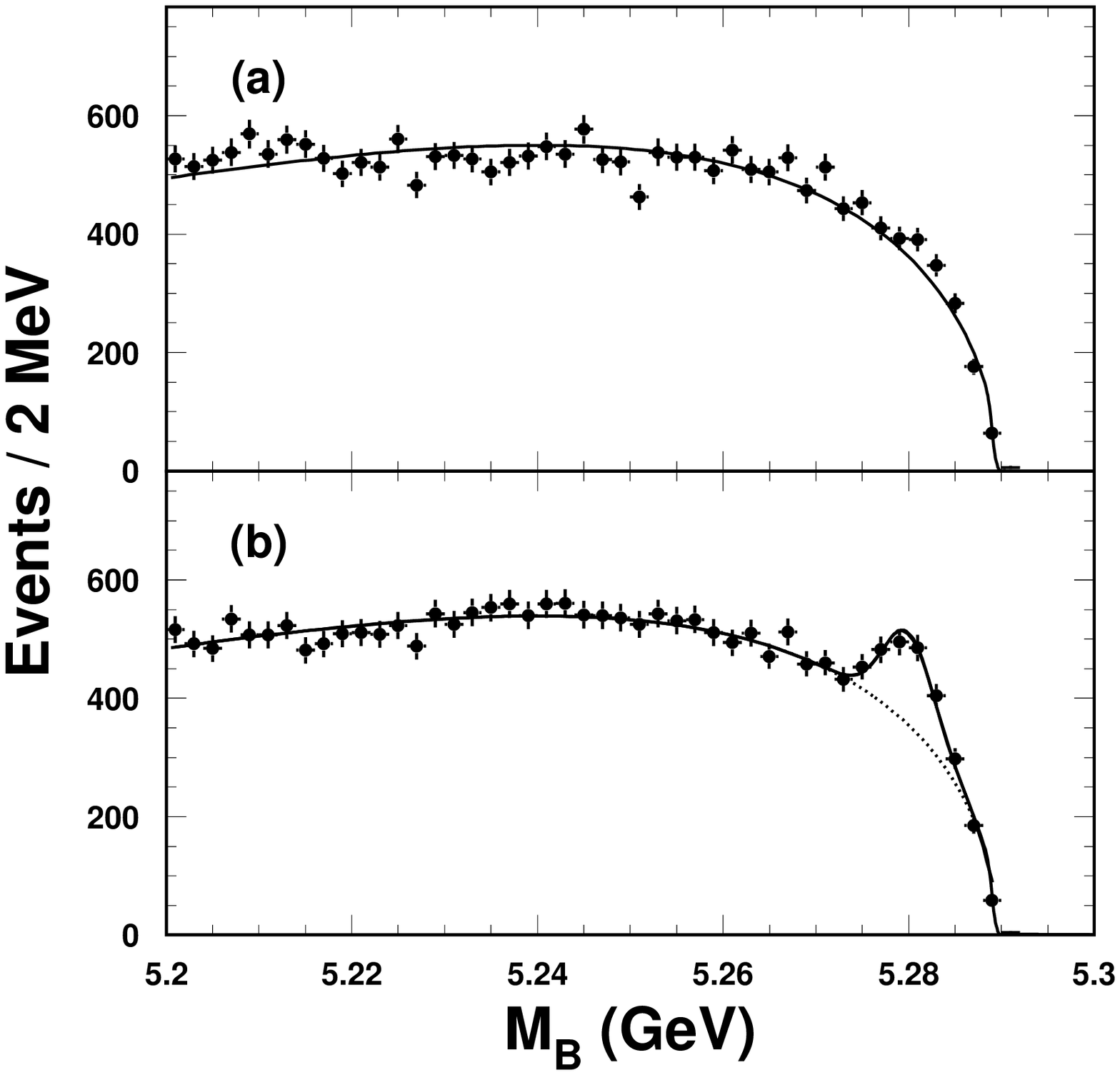,height=4in}\hspace{-0.9in}
\epsfig{figure=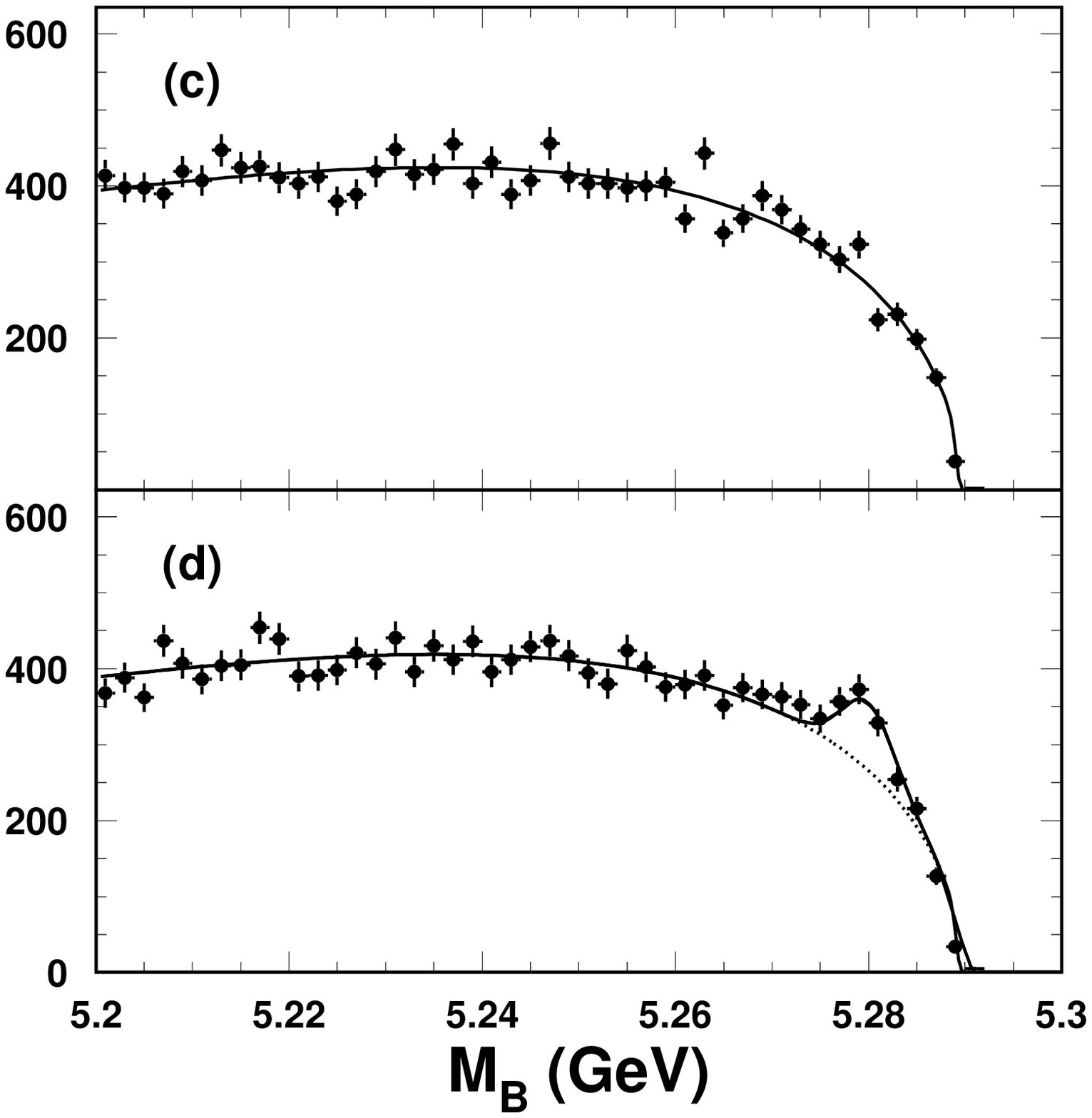,height=4in}}
\vspace{-1.0cm}
\caption{\label{bm_4pi_k2pi}The $B$ candidate mass spectra for
the final state $D^{*+}\pi^+\pi^-\pi^-\pi^o$, the left-side plots are
for $D^o\to K^-\pi^+\pi^o$ (a)  
$\Delta E$ sidebands, (b) for $\Delta E$ consistent with zero; the right-side
plots are for $D^o\to K^-\pi^+\pi^+\pi^-$ (c)  
$\Delta E$  sidebands, (d) for $\Delta E$ consistent with zero. The curves 
in the top plots, (a) and (c), are
fits to the background distribution described in the text, while in the 
bottom plots, 
(b) and (d), the
shapes from (a) and (c) are used with the normalization allowed to float and 
a signal Gaussian of width 2.7 MeV is added.
}
\end{figure}

 We next view the $M_B$ distribution for events having  $\Delta E$ 
within 2$\sigma$ around zero in
Fig.~\ref{bm_4pi_kpi}(b). This distribution is fit with a Gaussian 
signal 
function of width 2.7 MeV and the background function found above 
whose 
normalization is allowed to vary. We find 358$\pm$29 events in the 
signal peak.
 
We repeat this procedure for the other two $D^o$ decay modes. The 
$M_B$ spectrum for $\Delta E$ sidebands and signal region is shown 
in 
Fig.~\ref{bm_4pi_k2pi}. 
The $\Delta E$ 
resolution
is 22 MeV in the $K^-\pi^+\pi^o$ mode and 18 MeV in the 
$K^-\pi^+\pi^+\pi^-$ mode.
Signal to background ratios are worse in these two modes, but the
significance is still large. The number of signal events in each mode 
are shown in 
Table~\ref{table:4pievents}.

\begin{table}[hbt]
\begin{center}
\caption{Event numbers for the $D^{*+}\pi^+\pi^-\pi^-\pi^o$ final 
state}
\label{table:4pievents}
\begin{tabular}{lc}\hline\hline
$D^o$ Decay Mode & Fitted \# of events (\%)\\\hline
$K^-\pi^+$&               358$\pm$29       \\
$K^-\pi^+\pi^o$ &          543$\pm$49            \\
$K^-\pi^+\pi^+\pi^-$ &       329$\pm$41               \\\hline
\end{tabular}
\end{center}
\end{table}

\begin{figure}[H]
\centerline{\epsfig{figure=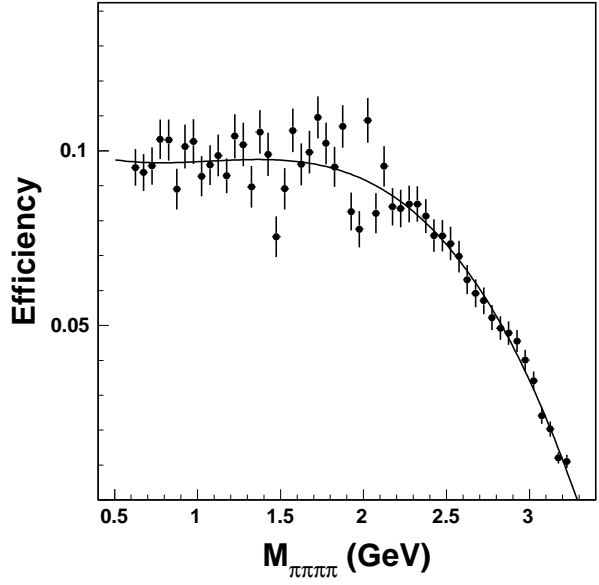,height=3.in}}
\vspace{0.3cm}
\caption{ \label{beff_ds4pi_kpi}The efficiency for the final state
$D^{*+}\pi^+\pi^-\pi^-\pi^o$, with $D^o\to K^-\pi^+$.
}
\end{figure}

\begin{figure}[htb]
\centerline{\epsfig{figure=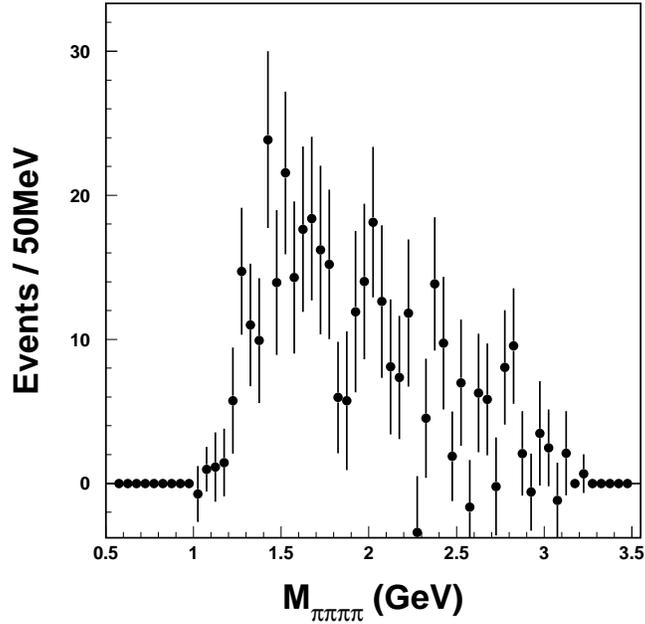,height=4in}}
\vspace{-1.0cm}
\caption{ \label{m4pi_ds4pi_kpi}The invariant mass spectra of 
$\pi^+\pi^-\pi^-\pi^o$ for the final state
$D^{*+}\pi^+\pi^-\pi^-\pi^o$, with $D^o\to K^-\pi^+$, found by 
fitting the
$B$ yield in bins of 4$\pi$ mass.
}
\end{figure}

We choose to determine the branching fraction using only the 
$D^o\to K^-\pi^+$ 
decay mode because of the relatively large backgrounds in the 
other modes and
the decreased systematic error due to having fewer particles in 
the final 
state.
In order to find the branching ratio we use the Monte Carlo 
generated efficiency, shown in Fig.~\ref{beff_ds4pi_kpi} as a 
function of $(4\pi)^-$ mass. The efficiency falls off at larger 
$(4\pi)^-$ masses because the detection
of the slow $\pi^+$ from the $D^{*+}$ decay becomes increasingly 
difficult.
Since the efficiency varies with mass we need to determine the 
$(4\pi)^-$ mass spectrum.
To rid ourselves of the problem of the background shape, we fit 
the $B$ 
candidate mass spectrum in 50 MeV bins of $(4\pi)^-$ mass. (The 
mass 
resolution
is approximately 12 MeV.) 
The resulting $(4\pi)^-$ mass 
spectrum is shown in Fig.~\ref{m4pi_ds4pi_kpi}. There are 
indications of a low mass structure around
1.4 GeV. 
This will be investigated further in this paper.

We find
\begin{equation}
{\cal B}(\overline{B}^o\to D^{*+}\pi^+\pi^+\pi^-\pi^o)
=(1.72\pm 0.14\pm 0.24)\%~~~.
\end{equation}

The systematic error arises mainly from our lack of knowledge 
about the 
tracking and $\pi^o$ efficiencies. We assign errors of $\pm$2.2\% 
on the 
efficiency of each charged track, $\pm$5\% for the slow pion from 
the $D^{*+}$, and 
$\pm$5.4\% for the $\pi^o$. The error due to the 
background shape is evaluated 
in three ways. First of all, we change the background shape by 
varying the
fitted parameters by 1$\sigma$. This results in a change of 
$\pm$3\%. Secondly,
we allow the shape, $p_2$, to vary (the normalization, $p_1$, was 
already
allowed to vary). This results in 3.8\% increase in the number of 
events.
Finally, we choose a different background function
\begin{equation}
back'(r)=p_1 r\sqrt{1-r^2}\left(1+p_2 r + p_3 r^2 +p_4 r^3 
\right)~~~,
\end{equation}
and repeat the fitting procedure. This results in a 3.7\% 
decrease in the
number of events. Taking a conservative estimate of the systematic 
error due to
the background shape we arrive at $\pm$3.8\%. We use the current 
particle data group
values for the relevant $D^{*+}$ and $D^o$ branching ratios of
(68.3$\pm$1.4)\% ($D^{*+}\to\pi^+ D^o$) and 
(3.85$\pm$0.09)\% ($D^o\to K^-\pi^+$), respectively \cite{PDG}. 
The relative
errors, 2.0\% for the $D^{*+}$ branching ratio and 2.3\% for the 
$D^o$ are
added in quadrature to the other sources of systematic error, 
yielding a
total systematic error of 10\%.

\begin{figure}[htb]
\centerline{\epsfig{figure=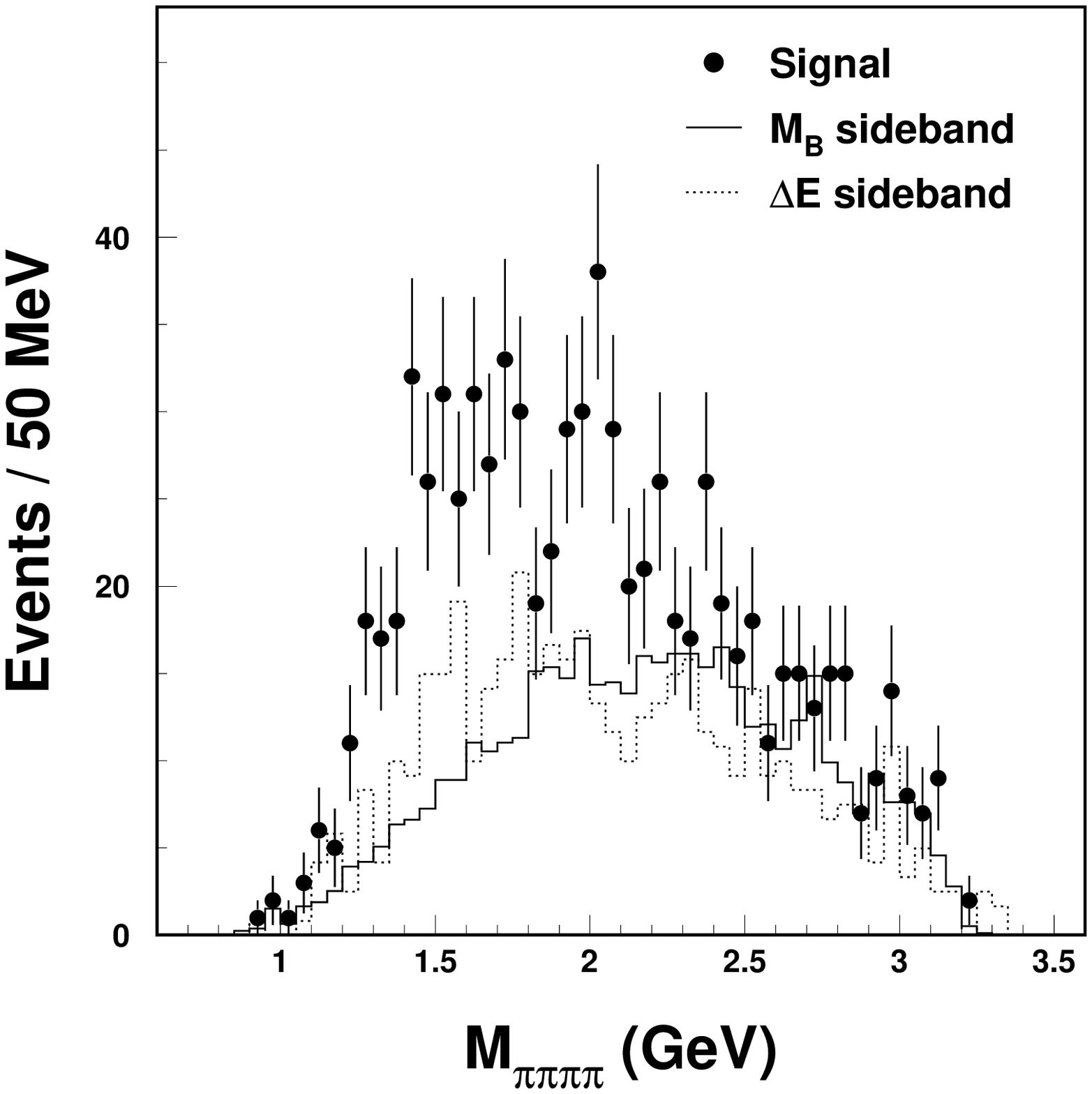,height=3.6in}
\epsfig{figure=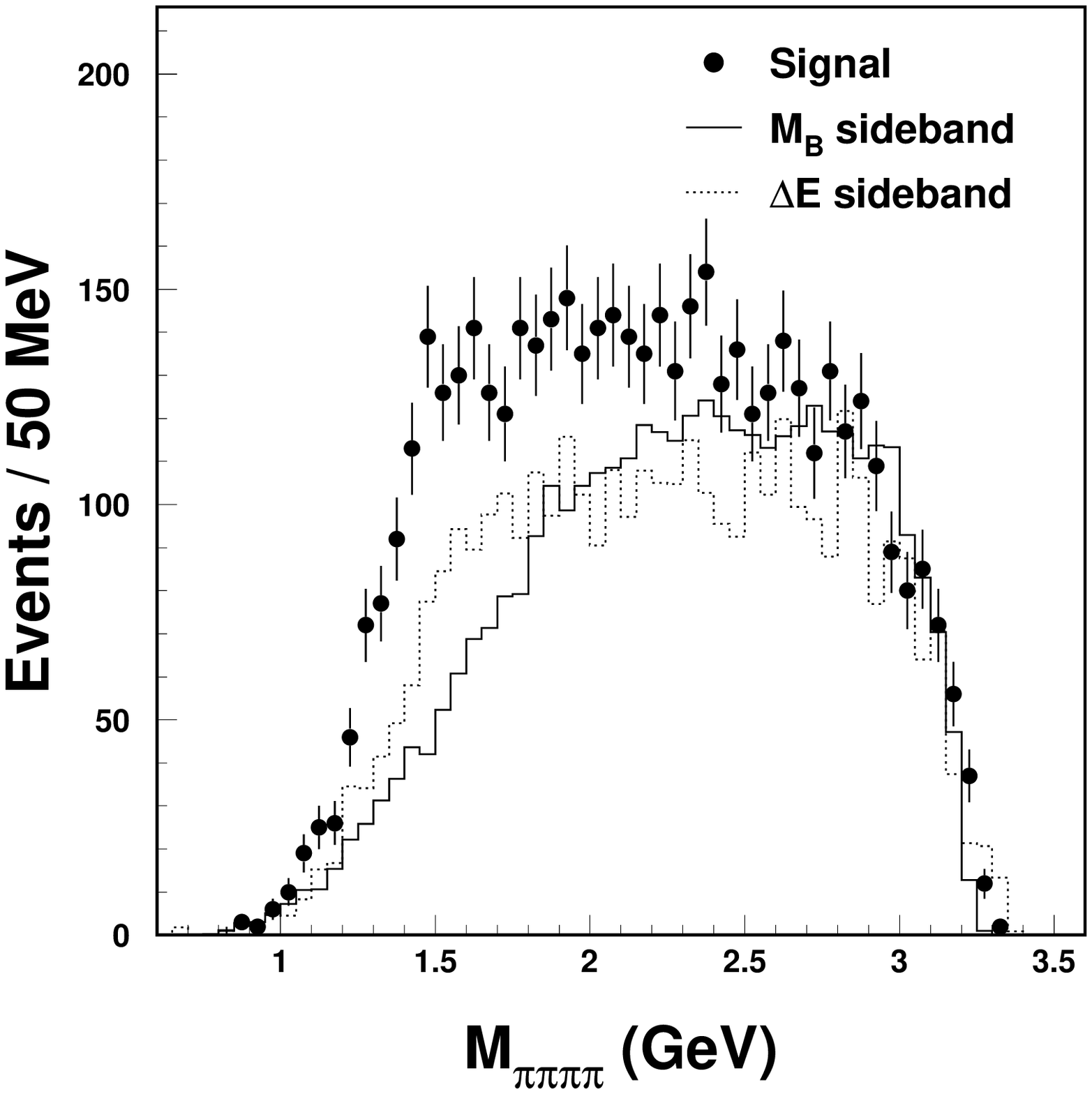,height=3.6in}}
\centerline{\epsfig{figure=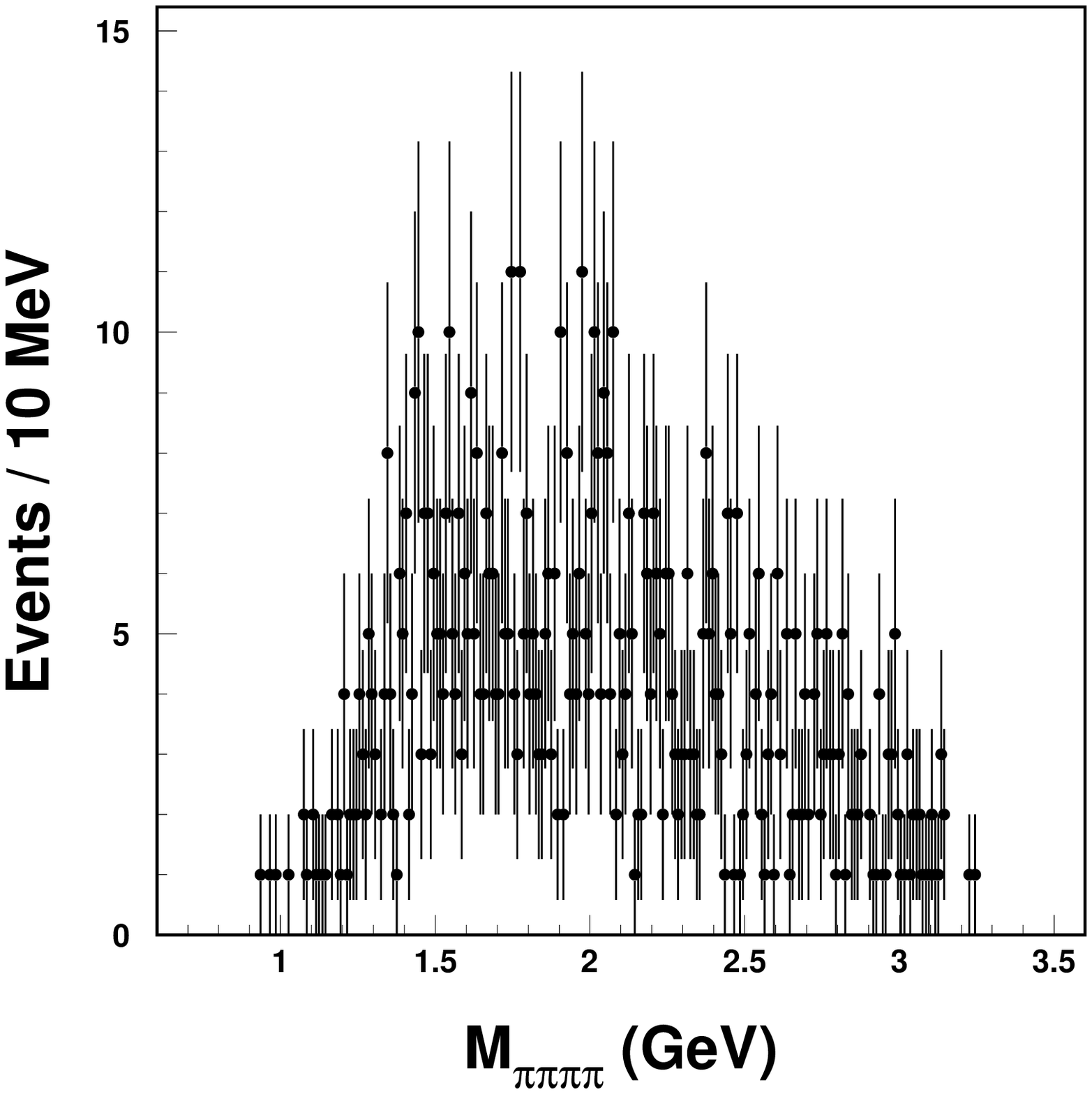,height=3.6in}
\epsfig{figure=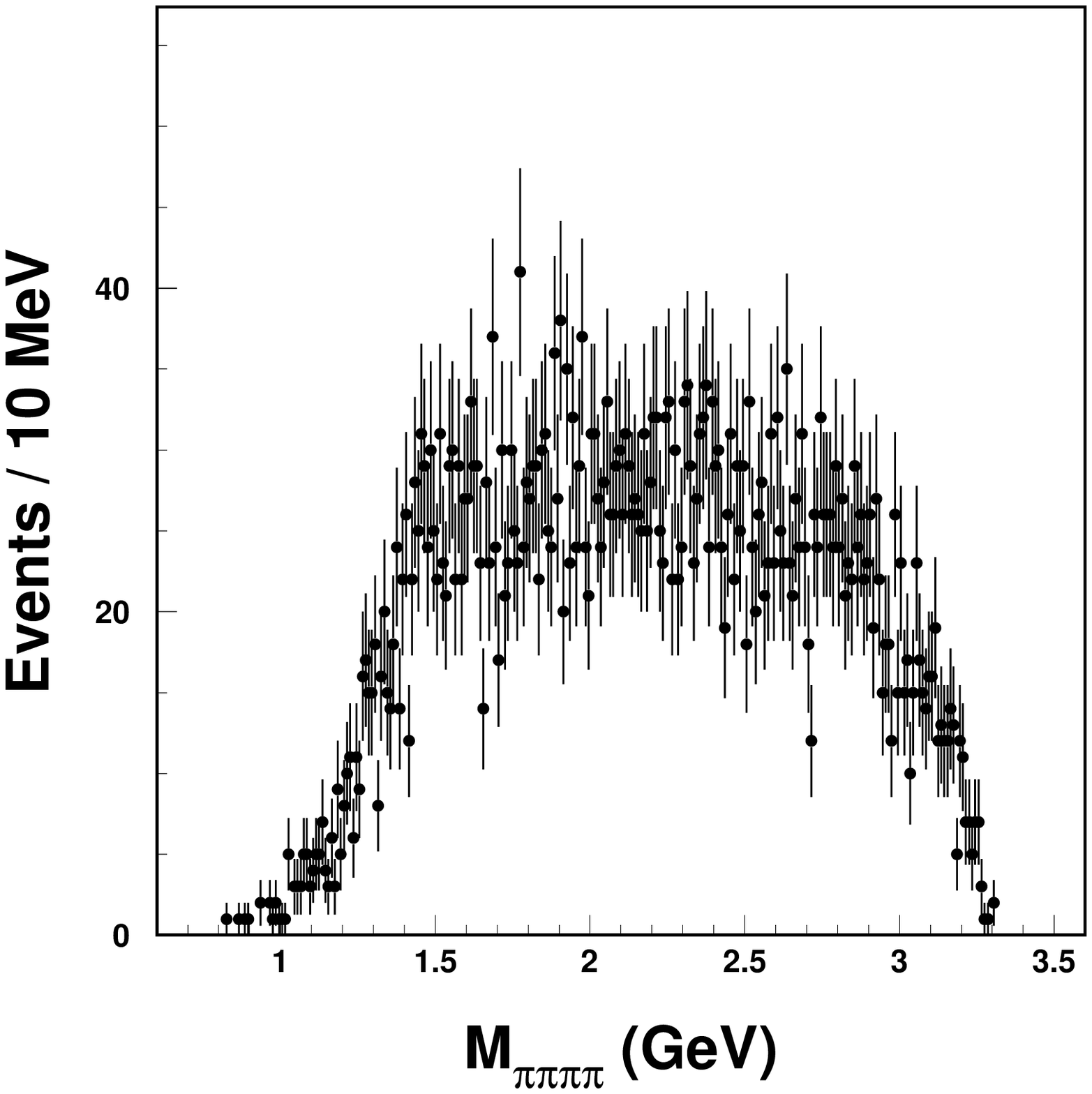,height=3.6in}}
\vspace{-1.0cm}
\caption{\label{m4pi_c_all}The invariant mass spectra of 
$\pi^+\pi^-\pi^-\pi^o$ for the final state
$D^{*+}\pi^+\pi^-\pi^-\pi^o$, with $D^o\to K^-\pi^+$ (upper left), 
and the sum of all three $D^o$ decay modes (upper right). Events are selected by being
within $2\sigma$ of the $B$ mass. The solid histogram is the background
estimate from the $M_B$ lower sideband and the dashed histogram is from the
$\Delta E$ sidebands; both are normalized to the fitted number of background
events. The same distributions in smaller bins (lower plots).
}
\end{figure}

We wish to search for narrow structures. However, we cannot fit the $B$ mass
spectrum in small $(4\pi)^-$ mass intervals due to a lack of statistics.
Thus we plot the  $(4\pi)^-$ mass for events
 in the $M_B$ peak for the $D^o\to K^-\pi^+$ mode and the sum of all three modes in 
 Fig.~\ref{m4pi_c_all}. We also plot two background samples: events at lower
$M_B$ (5.203 - 5.257 GeV) and those in the $\Delta E$ sideband separately.
\afterpage{\clearpage}
First we view the plots in the canonical 50 MeV bins. Both background
distributions give a consistent if somewhat different estimates of the
background shape. (Each background distribution has been normalized
to the absolute number of background events as determined by the
fit to the $M_B$ distribution.) In any case no prominent narrow
structures appear in the histograms for the 10 MeV binning.

\section{The $\overline{B}^o\to D^{*+}\omega\pi^-$ Reaction}
\label{sec:Dspomegapi}
\begin{figure}[H]
\vspace{-.5cm}
\centerline{\epsfig{figure=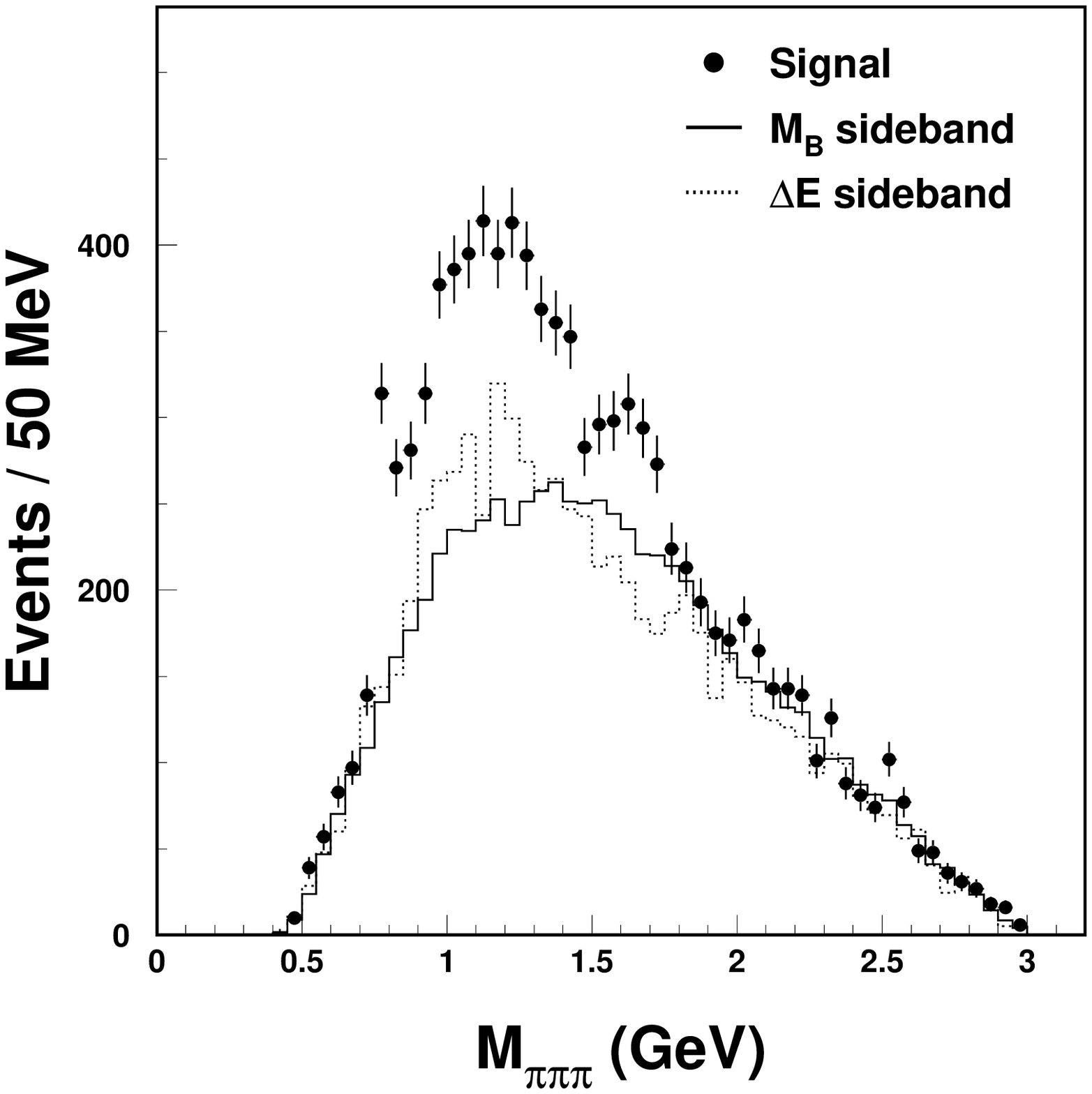,height=4in}}
\vspace{-1.0cm}
\caption{ \label{m3pi_c_all}The invariant mass spectra of 
$\pi^+\pi^-\pi^o$ for the final state
$D^{*+}\pi^+\pi^-\pi^-\pi^o$ for all three $D^o$ decay modes.
The solid histogram is the background
estimate from the $M_B$ lower sideband and the dashed histogram is 
from the
$\Delta E$ sidebands; both are normalized to the fitted number of 
background
events. 
}
\end{figure}

To investigate the composition of the $(4\pi)^-$ final state,
we now  investigate the $\pi^+\pi^-\pi^o$ mass spectrum for the
events in the $B$ peak. All three $D^o$ decay modes are used. 
We show the $\pi^+\pi^-\pi^o$ invariant mass distribution for 
events
in the $B$ mass peak in Fig.~\ref{m3pi_c_all} (there are two 
combinations
per event). A clear signal is visible at the
 $\omega$. The histograms on the figure are for events either in the lower 
 $M_B$ range, from 5.203 GeV to 5.257 GeV, or in the previously defined
$\Delta E$ sidebands; no $\omega$ signal is visible.

The purity of the $\omega$ sample can be further improved by 
restricting candidates to certain regions of the 
Dalitz plot of the decay products. We define a cut on the Dalitz 
plot as follows.
Let $T_0$, $T_+$ and $T_-$ be the kinetic energies of the pions, 
and $Q$ be the difference between the $\omega$ mass, $M_{\omega}$, 
and the mass of the 3 pions. We define two orthogonal coordinates 
$X$ and $Y$, where
\begin{eqnarray}
        X &= &3T_0/Q - 1       \\
        Y &= &\sqrt{3}(T_+ - T_-)/Q~~~.
\end{eqnarray}
The  kinematic limit that defines the Dalitz plot boundary is 
defined as
\begin{equation}
        Y_{boundary}^2 = {1\over 3} 
(X_{boundary}+1)(X_{boundary}+1+a) (1+b/(X_{boundary}+1-c))
\end{equation}
        where $a = 6m_0 / Q$,   $b = 6m^2 / (M_{\omega}Q)$, 
              $c = 3(M_{\omega}-m_0)^2/(2M_{\omega}Q)$, $m$ is the 
mass of a charged pion and $m_0$ the mass of the neutral pion.

For any set of three pion kinetic energies, we define a variable 
$r$, properly scaled to the kinematic limit as
\begin{equation}
r=\sqrt{{X^2+Y^2}\over{X_{boundary}^2+Y_{boundary}^2}}~~~,
\end{equation}
where the boundary values are found by following the radial vector 
from (0,0) through $(X,Y)$.

\begin{figure}[H]
\vspace{-0.3cm}
\centerline{\epsfig{figure=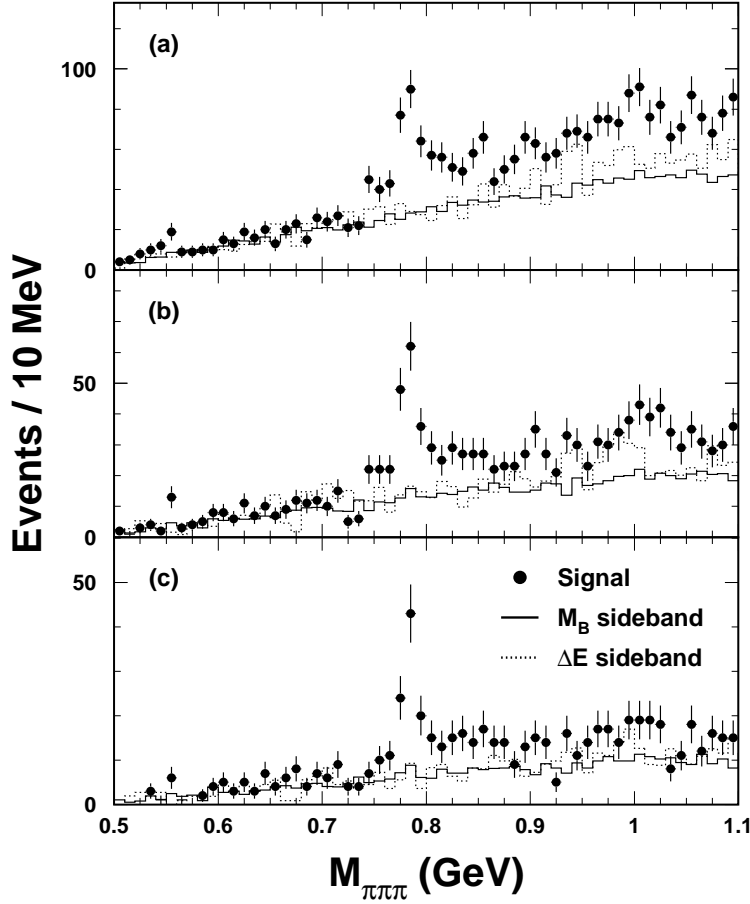,height=5.6in}}
\vspace{-1.0cm}
\caption{ \label{m3_dal_all}The invariant mass spectra of 
$\pi^+\pi^-\pi^o$ for the final state
$D^{*+}\pi^+\pi^-\pi^-\pi^o$ for all three $D^o$ decay modes for
three selections on $r$:  (a) 1, (b) 0.7 and (c) 0.5.
The solid histogram is the background
estimate from the $M_B$ lower sideband and the dashed histogram is 
from the
$\Delta E$ sidebands; both are normalized to the fitted number of 
background events.}
\end{figure}

For events in the $B$ mass peak we show in Fig.~\ref{m3_dal_all} 
the
$\pi^+\pi^-\pi^o$ invariant mass for three different cuts on $r$. 
The $\omega$ signal is purified by using a selection on $r$.
 
\begin{figure}[htb]
\centerline{\epsfig{figure=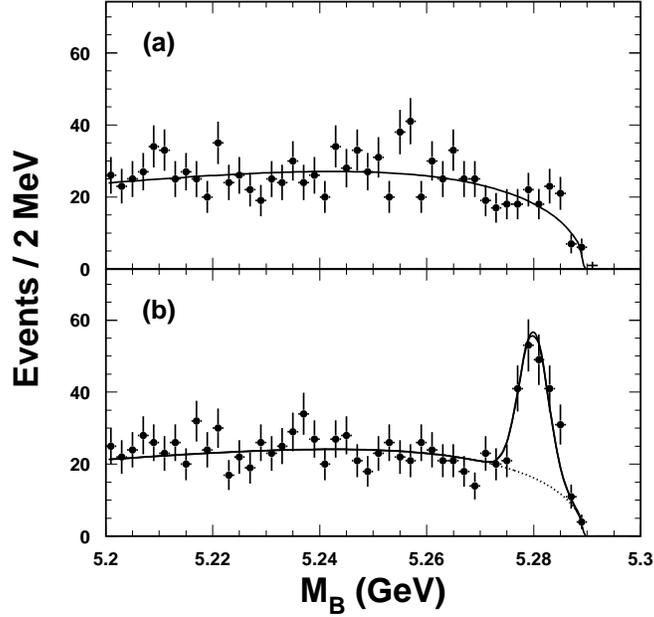,height=4in}}
\vspace{-1.0cm}
\caption{ \label{bm_4pi_all_4}The $M_B$ spectra for  
$D^{*+}\omega\pi^-$ for all three $D^o$ decay modes.
(a) $\Delta E$ sidebands and (b) $\Delta E$ around zero.
 }
\end{figure}

For further analysis we select $\omega$ candidates within the the 
$\pi^+\pi^-\pi^o$ mass window of 782$\pm$20 MeV with $r<0.7$. We 
also abandon
the $\chi^2$ cut as background is less of a problem. In 
Fig.~\ref{bm_4pi_all_4}
 we show the
$B$ candidate mass distribution for the $D^{*+}\omega\pi^-$ final 
state summing
over all three $D^o$ decay modes. (The signal is fit with the same
prescription as before.) There are 136$\pm$15 events in the peak.
\begin{figure}[htb]
\centerline{\epsfig{figure=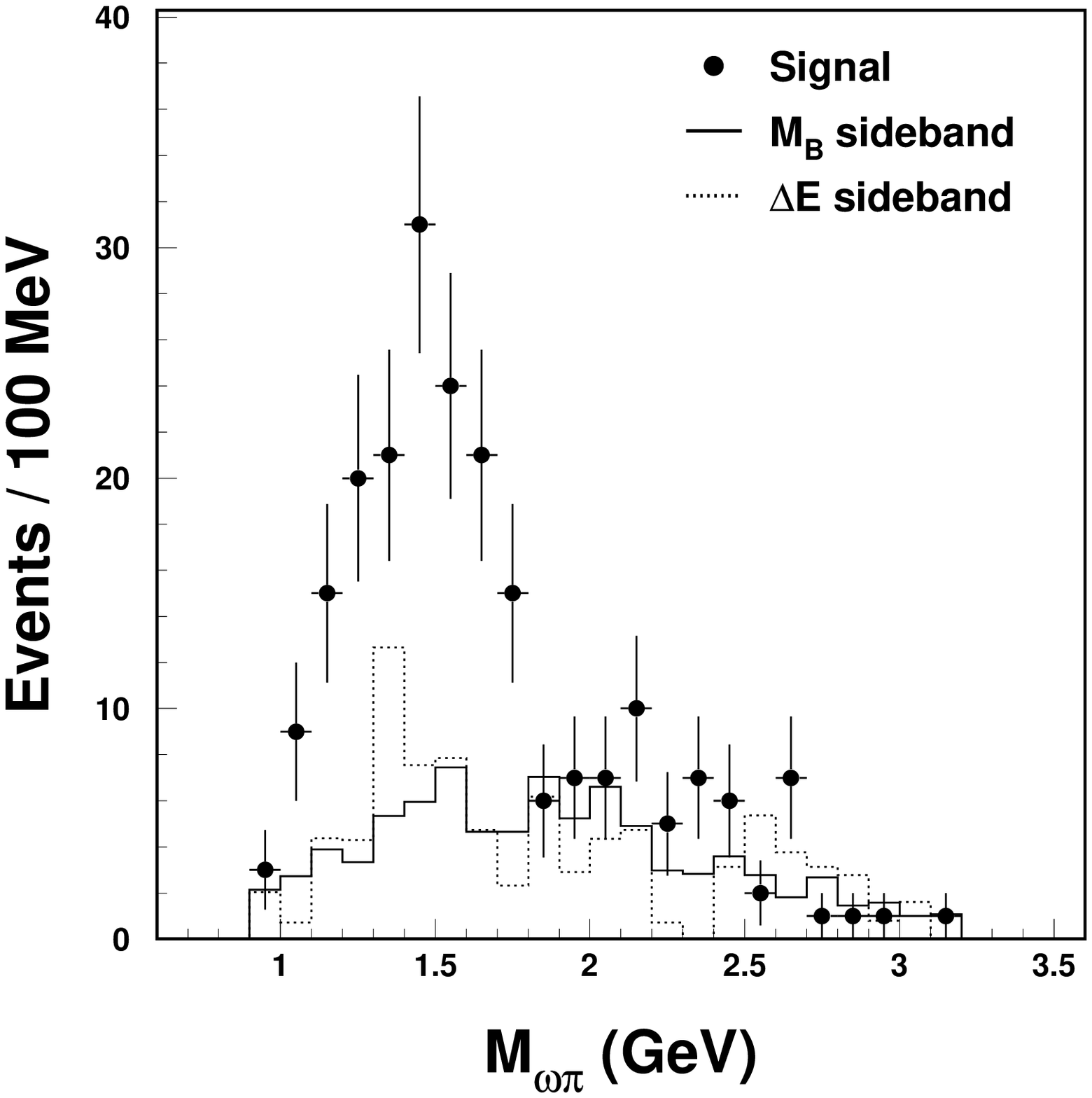,height=3.7in}
\epsfig{figure=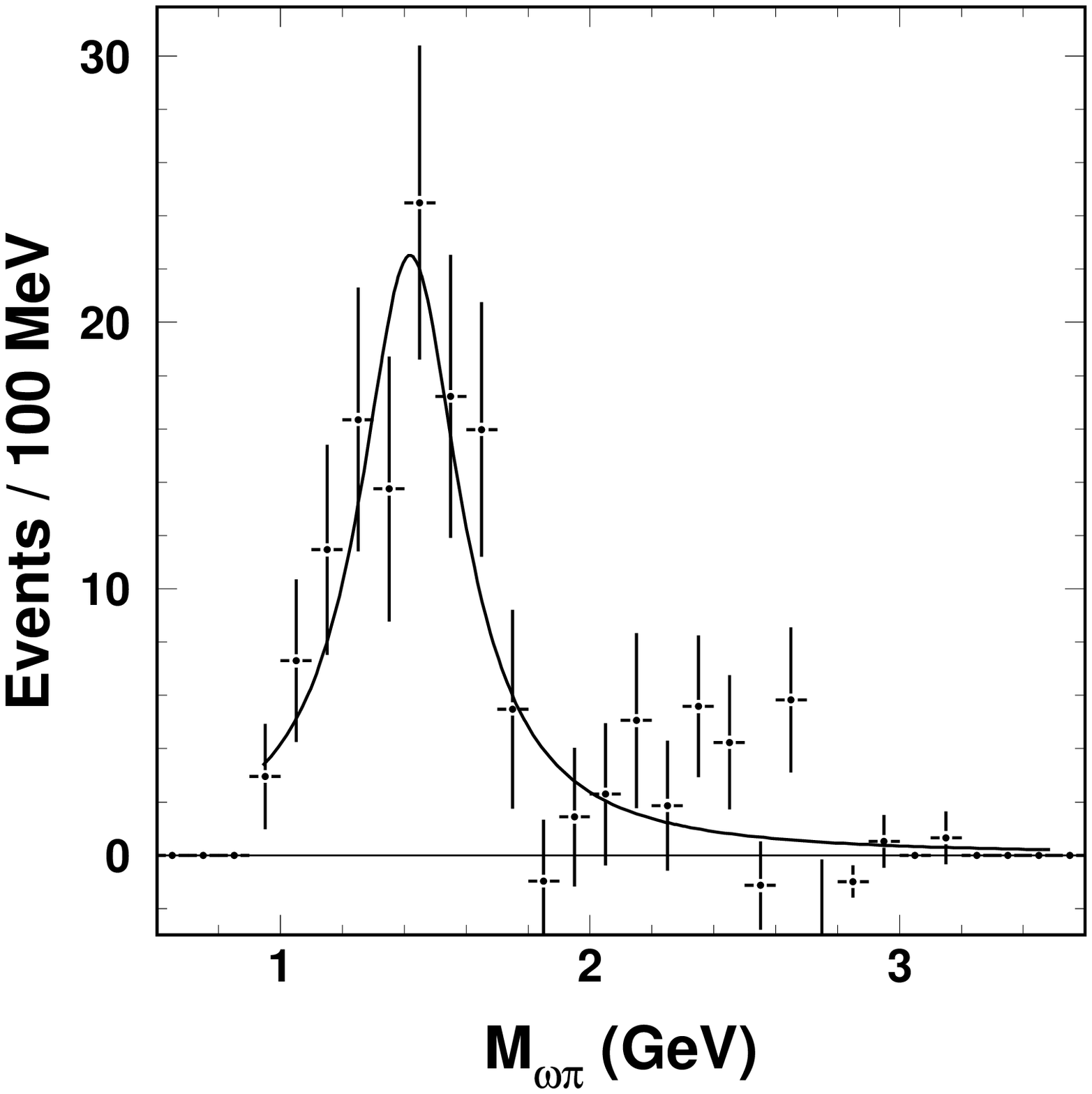,height=3.7in}}
\vspace{-1.0cm}
\caption{ \label{m4pi_c4_all}The invariant mass spectra of 
$\omega\pi^-$ for the final state
$D^{*+}\pi^+\pi^-\pi^-\pi^o$ for all three $D^o$ decay modes.
(left) The solid histogram is the background
estimate from the $M_B$ lower sideband and the dashed histogram is 
from the
$\Delta E$ sidebands; both are normalized to the fitted number of 
background
events. (right) The mass spectrum determined from fitting the 
$M_B$ distribution
and fit to a Breit-Wigner function. 
}
\end{figure}	
\afterpage{\clearpage}

In Fig.~\ref{m4pi_c4_all} we show
the $\omega\pi^-$ mass spectrum in the left-side plot. The solid 
histogram shows events from the 
lower $M_B$ sideband region suitably normalized. The dotted 
histogram shows 
the background estimate from the $\Delta E$ sidebands, again 
normalized. 
In the signal distribution there is a wide structure around 
1.4 GeV, that is inconsistent with background. We re-determine the 
$\omega\pi^-$
mass distribution by fitting the $M_B$ distribution in bins of 
$\omega\pi^-$
mass, and this is shown on the right-side.  Fitting to a Breit-Wigner function,
we find a peak value of 1416$\pm$37 MeV and a width of 402$\pm$47 
MeV. These numbers change to 1432$\pm$37 MeV and 376$\pm$47 MeV, 
respectively, after applying a correction for 
the variation of efficiency with mass.

Knowing the $\omega\pi^-$ mass dependence of the efficiency we evaluate the 
branching fraction:
\begin{equation}
{\cal B}(\overline{B}^o\to D^{*+}\omega\pi^-)=(0.29 \pm 0.03 \pm 
0.04)\%~~~.
\end{equation}

We tentatively label the state at 1432 MeV the $A^-$ and 
investigate its
properties later. The $\omega\pi^-$ comprises about 17\% of the 
$(4\pi)^-$ final state.
All of the $\omega\pi^-$ final state is consistent with coming 
from $A^-$ decay.

\section{Observation of ${B}^-\to D^{*o}\pi^+\pi^-\pi^-\pi^o$}
\label{sec:Ds04pi}
\begin{figure}[bht]
\vspace{-1.5cm}
\centerline{\epsfig{figure=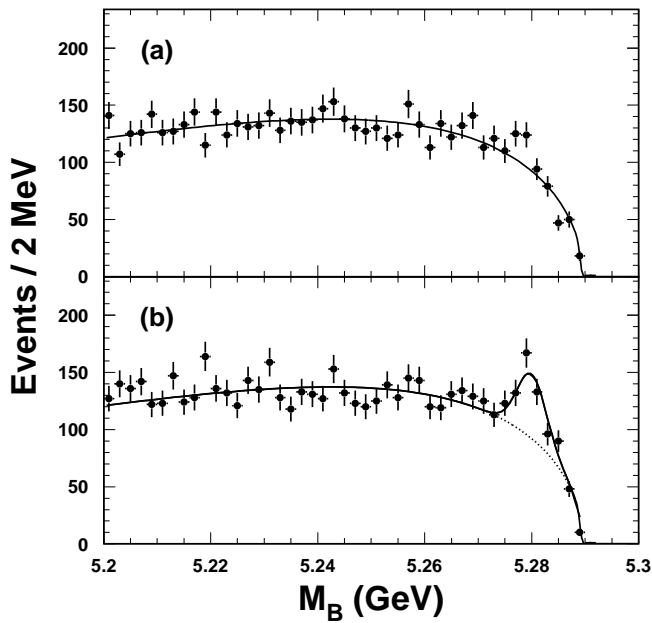,height=4in}}
\vspace{-1.cm}
\caption{ \label{bm_4pi_0kpi}The $B$ candidate mass spectra for 
the final state
$D^{*o}\pi^+\pi^-\pi^-\pi^o$, with $D^o\to K^-\pi^+$ (a) for 
$\Delta E$
sidebands and (b) for $\Delta E$ consistent with zero. The curve 
in (a) is a
fit to the background distribution described in the text, while in 
(b) the
shape from (a) is used with the normalization allowed to float and 
a signal
Gaussian of width 2.7 MeV is added.
}
\end{figure}

We proceed in the same manner as for the $\overline{B}^o$ reaction 
with the exception that we use the $D^{*o}\to \pi^o D^o$ decay 
mode and 
restrict ourselves to the $D^o\to K^-\pi^+$ decay mode only due to 
large backgrounds
in the other modes. The $\chi^2$ is calculated according to 
equation~\ref{eq:chisq} and we use a cut value of 8.
The $M_B$ distributions for $\Delta E$ sidebands and signal data are 
shown in 
Fig.~\ref{bm_4pi_0kpi} for the $D^o\to K^-\pi^+$ decay mode. We 
see a signal 
of 195$\pm$26 events 
yielding a branching fraction of 
\begin{equation}
{\cal{B}}({B}^-\to D^{*o}\pi^+\pi^-\pi^-\pi^o) = (1.80\pm 0.24\pm 
0.25) \%~~~~.
\end{equation}

\begin{figure}[thb]
\centerline{\epsfig{figure=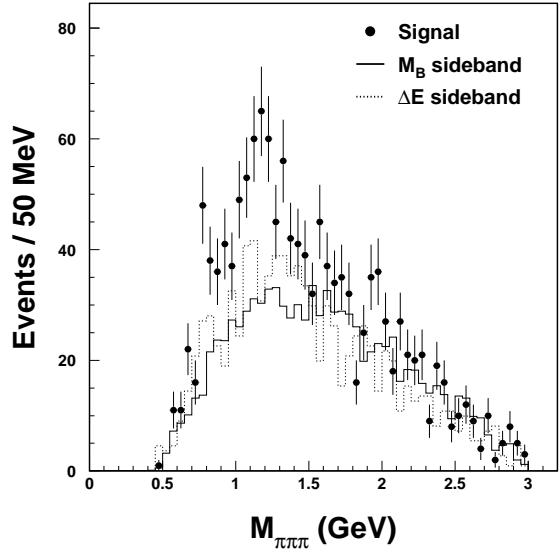,height=3.5in}}
\vspace{-1.0cm}
\caption{ \label{m3pi_c_0kpi}The invariant mass spectra of 
$\pi^+\pi^-\pi^o$ for the final state
$D^{*o}\pi^+\pi^-\pi^-\pi^o$ for the $D^o\to K^-\pi^+$ decay 
mode.
The solid histogram is the background
estimate from the $M_B$ lower sideband and the dashed histogram is 
from the
$\Delta E$ sidebands; both are normalized to the fitted number of 
background
events. There are two combinations per event. 
}
\end{figure}

\begin{figure}[htb]
\centerline{\epsfig{figure=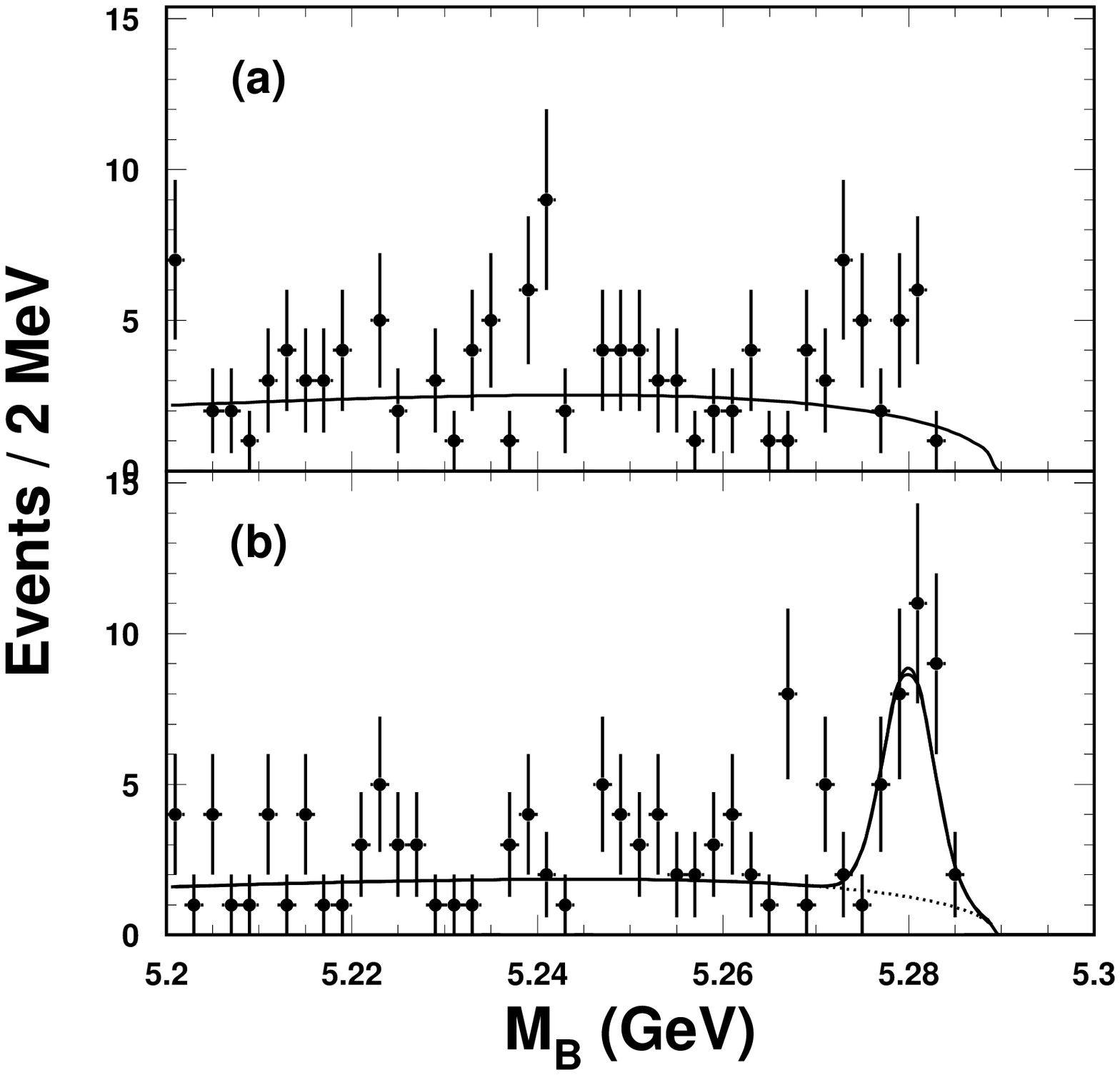,height=4in}}
\vspace{-1.0cm}
\caption{ \label{bm_4pi_0kpi_4}The $M_B$ spectra for  
$D^{*o}\omega\pi^-$ for the $D^o\to K^-\pi^+$ decay mode.
(a) $\Delta E$ sidebands and (b) $\Delta E$ around zero.
 }
\end{figure}

The $\pi^+\pi^-\pi^o$ mass spectrum shown in 
Fig.~\ref{m3pi_c_0kpi} shows the presence of an $\omega$.
Selecting on the presence of an $\omega$ with $r<0.7$ we show the 
sideband and signal plots in Fig.~\ref{bm_4pi_0kpi_4}.
(Here we do not use the previously defined $\chi^2$ cut.)
\afterpage{\clearpage}
The branching ratio is
\begin{equation}
{\cal{B}}({B}^-\to D^{*o}\omega\pi^-) = (0.45\pm 0.10 \pm 0.07) 
\%~~~~.
\end{equation}

\begin{figure}[htb]
\centerline{\epsfig{figure=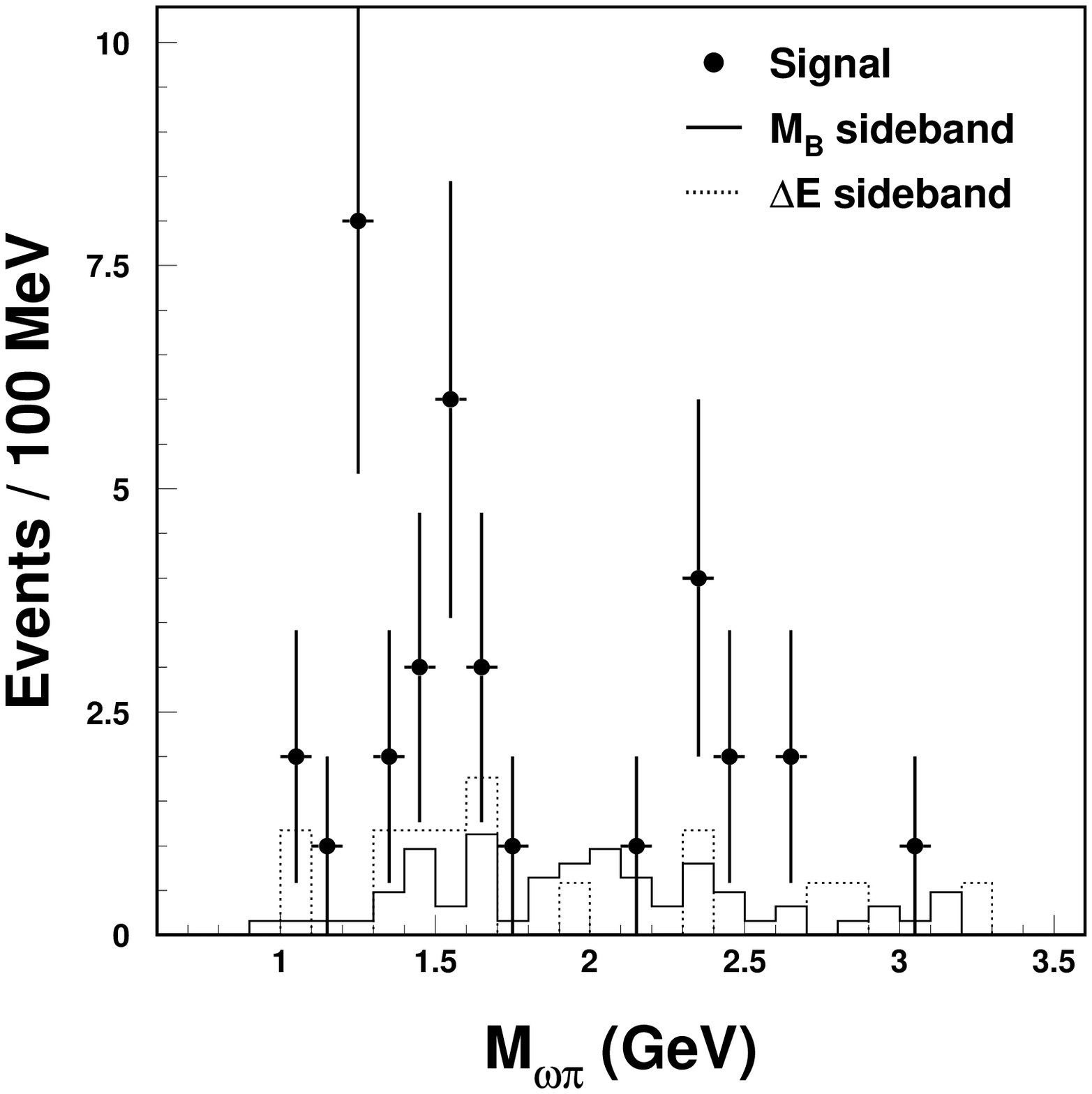,height=3.7in}
\epsfig{figure=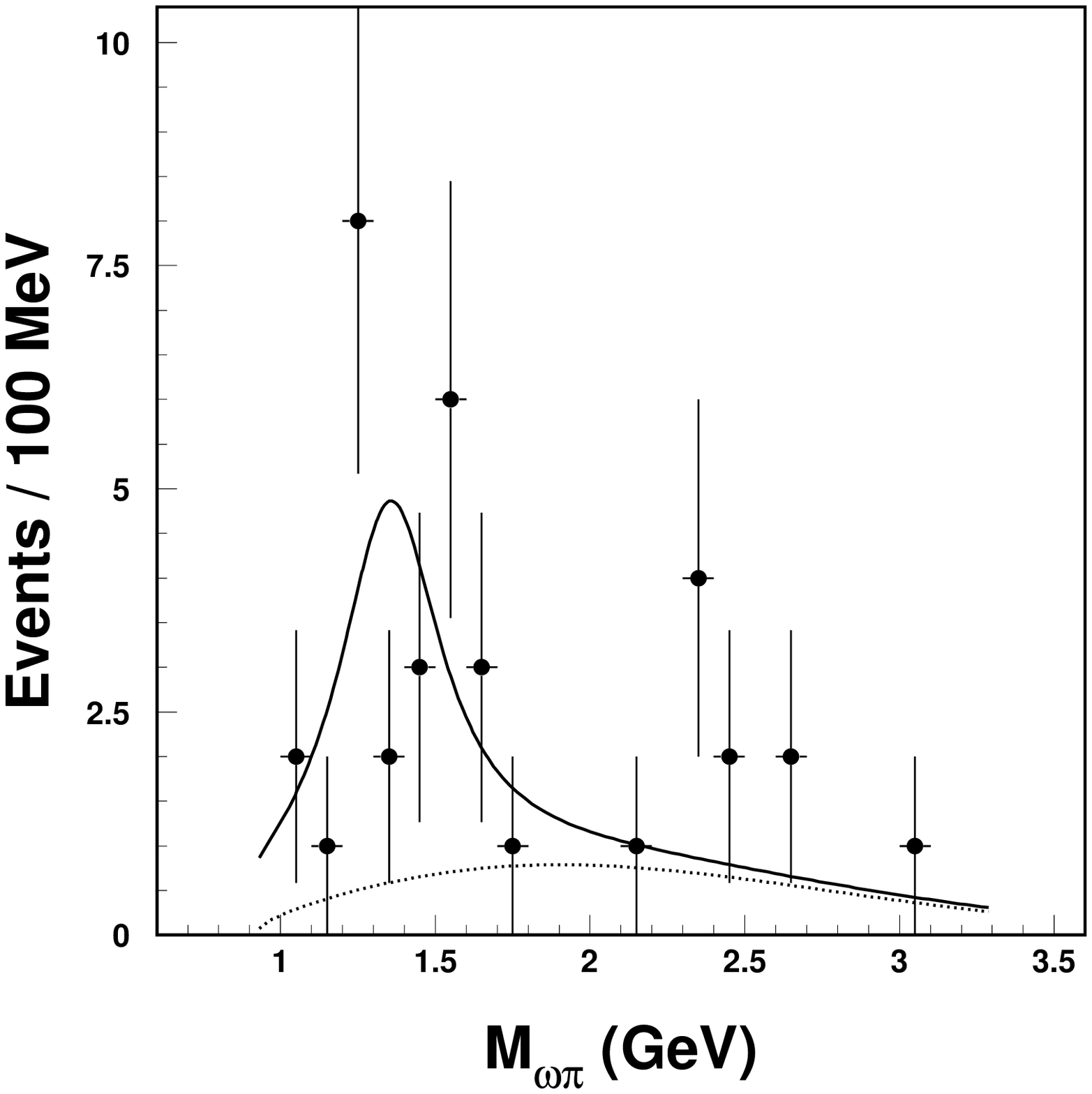,height=3.7in}}
\vspace{-1.0cm}
\caption{ \label{m4pi_c4_0kpi}The invariant mass spectra of 
$\omega\pi^-$ for the final state
$D^{*o}\pi^+\pi^-\pi^-\pi^o$ for the $D^o\to K^-\pi^+$ decay mode.
(left) The solid histogram is the background
estimate from the $M_B$ lower sideband and the dashed histogram is 
from the
$\Delta E$ sidebands; both are normalized to the fitted number of 
background
events. (right) The data fit to a Breit-Wigner signal and a smooth 
background
function. 
}
\end{figure}

In Fig.~\ref{m4pi_c4_0kpi} we show the $\omega\pi^-$
mass spectrum.
We see an enhancement at around 1.4 GeV as in the
neutral $B$ case. A fit to the data gives a mass of 1367$\pm$75 
MeV 
and width of 439$\pm$135 MeV, consistent within the large errors 
with
the $\overline{B}^o$ case. (We do not have enough statistics here 
to 
fit the $M_B$ distribution in bins of $\omega\pi^-$ mass.)
The $\omega\pi^-$ fraction of the 
$(4\pi)^-$ final 
state is 25\%, and all the $\omega\pi^-$ is consistent with coming 
from the
$A^-$.

\section{Analysis of $D^{*+}\omega\pi^-$ Decay Angular Distributions}
\label{sec:angular}

 The $A^-$ is produced along with a spin-1 $D^*$ from a spin-0 
$B$. 
If the $A^-$ is spin-0 the $D^*$ would be fully polarized in the 
$(J,~J_z)= (1,0)$ state.
If the $A^-$ were to be spin-1 any combinations of z-components 
would be
allowed. It is natural then to examine the helicity angle of the 
$D^{*+}$ by
viewing the cosine of the helicity angle of the $\pi^+$ with 
respect to the
$B$ in the $D^{*+}$ rest frame.

Another decay angle that can be examined is that of the 
$\omega\pi$ system.
If the $A^-$ is spin-0, the $\omega$ is polarized in the (1,0) 
state
and may be if the $A^-$ is spin-1. Here the helicity angle is 
defined as the
angle between the normal to the $\omega$ decay plane and the 
direction of the
$A^-$ in the $\omega$ rest frame. For a spin-0 $A^-$ the 
distribution will
be cosine-squared. Again full polarization is possible if the $A^-
$ is other 
than spin-0, but any distribution other than cosine-squared would 
demonstrate
that the spin is not equal to zero.

For this analysis we use all three $D^o$ final states for the 
$D^{*+}$ final 
state. To find the distributions we fit the number of events
in the $M_B$ candidate plot selected on different angle bins. The 
$\omega\pi$
mass is required to be between 1.1 and 1.9 GeV. This leaves 111$\pm$13 events.

\begin{figure}[bht]
\centerline{\epsfig{figure=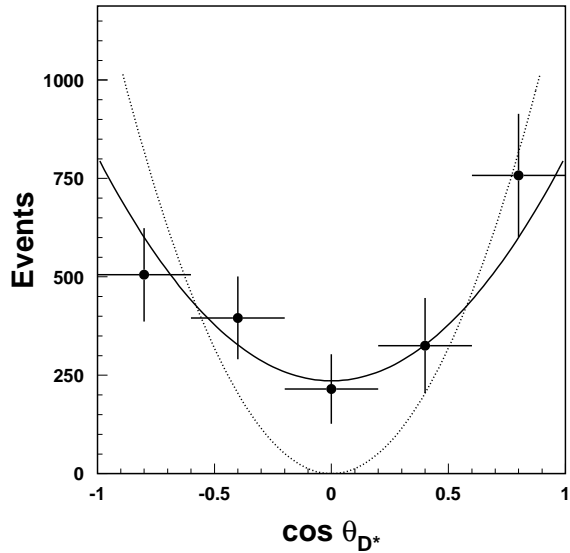,height=3.5in}}
\vspace{-1.0cm}
\caption{ \label{cosd} The cosine of the angle between the $D^o$ 
and the
$D^{*}$ flight direction in the $D^{*}$ rest frame for the 
$D^{*}A^-$ final
state (solid points) along with a fit (solid curve) allowing the
amount of longitudinal and traverse polarization to vary. The dotted curve
is the expectation for a spin-0 $A^-$.}
\end{figure}

In Fig.~\ref{cosd} we show the helicity angle distribution, 
$\cos\theta_{D^*}$ 
for the $D^{*}$ decay. The data have been corrected for acceptance.
 We also show the expectation for 
spin-0 from the Monte Carlo. The data have been fit for the fraction
of longitudinal polarization. We find 
\begin{equation}
{\Gamma_L \over \Gamma}=0.63\pm 0.09 ~~.
\end{equation}

\begin{figure}[htb]
\centerline{\epsfig{figure=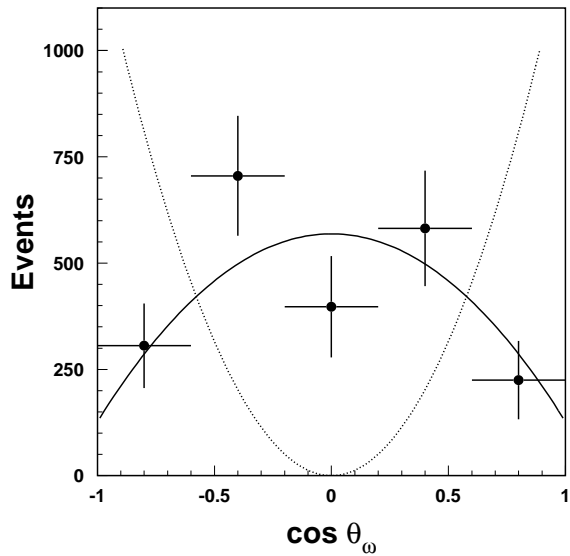,height=3.5in}}
\vspace{-0.4cm}
\caption{\label{cosom} The cosine of the angle between the normal 
to the
$\pi^+\pi^-\pi^o$ decay plane and the $\omega$ boost direction
 for the $D^{*}A^-$ final state (solid points) along with a fit (solid curve)
 allowing the
amount of longitudinal and traverse polarization to vary. The dotted
curve is the expectation for a spin-0 $A^-$.}
\end{figure}
The $\cos\theta_{D^*}$ distribution is not consistent with full 
polarization, yielding a $\chi^2$ of 17.7 for 5 degrees of freedom. 
The helicity angle
distribution for the $A^-\to\omega\pi^-$, $\cos\theta_{\omega}$, 
is shown on Fig.~\ref{cosom}. Furthermore 
the $\cos\theta_{\omega}$ distribution is quite inconsistent with 
a $\cos^2\theta_{\omega}$, yielding a $\chi^2$ of 109 for 5 degrees 
of freedom. Therefore, we rule out a spin-0 assignment for the $A^-$.

To determine the $J^P$ we need a more well defined final state. 
This is provided by analysis of $B\to D\omega\pi^-$ decays.

\section{Observation of $\overline{B}\to D\omega\pi^-$ Decays}
\label{sec:Domegapi}  
\subsection {$B$ candidate selection}  
  
Here we study the reactions $\overline{B}\to D\omega\pi^-$, with 
either a 
$D^o\to K^-\pi^+$ or $D^+\to K^-\pi^+\pi^+$ decay. Other $D^o$ or 
$D^+$ decays have substantially larger backgrounds. 
    
Although we are restricting our search to $\omega$'s, we define 
two $\pi^+\pi^-\pi^0$
samples. One within 20 MeV of the known $\omega$ mass (782 MeV) 
and the other in 
either low mass or high mass sideband defined as three $\pi$ mass 
either between
732 and 752 MeV or between 812 and 832 MeV. We also require a cut 
on the $\omega$
Dalitz plot of $r < 0.7.$

To reduce backgrounds we define 
\begin{equation}\label{eq:chisqD} 
\chi_b^2=  
\left({{\Delta M_{D}}\over {\sigma(\Delta M_{D})}}\right)^2 +  
\left({{\Delta M_{\omega}}\over {\sigma(\Delta 
M_{\omega})}}\right)^2 +  
\left({{\Delta M_{\pi^o}}\over {\sigma(\Delta 
M_{\pi^o})}}\right)^2~~~, 
\end{equation} 
where $\Delta M_{D}$ is the invariant candidate $D^o$ mass minus 
the known $D^o$
mass, $\Delta M_{\omega}$ is the invariant candidate $\omega$ mass 
minus the known
$\omega$ mass, and $\Delta M_{\pi^o}$ is the measured 
$\gamma\gamma$ invariant
mass minus the known $\pi^o$ mass. The $\sigma$'s are the 
measurement errors.
We select candidate events requiring that  $\chi^2_b$ is
$<$ 12 for the $K\pi$ mode and $<$6 for the $K\pi\pi$. 

\subsection{$B^-\to D^o\omega\pi^-$ Signal}

\begin{figure}[htb] 
\centerline{\epsfig{figure=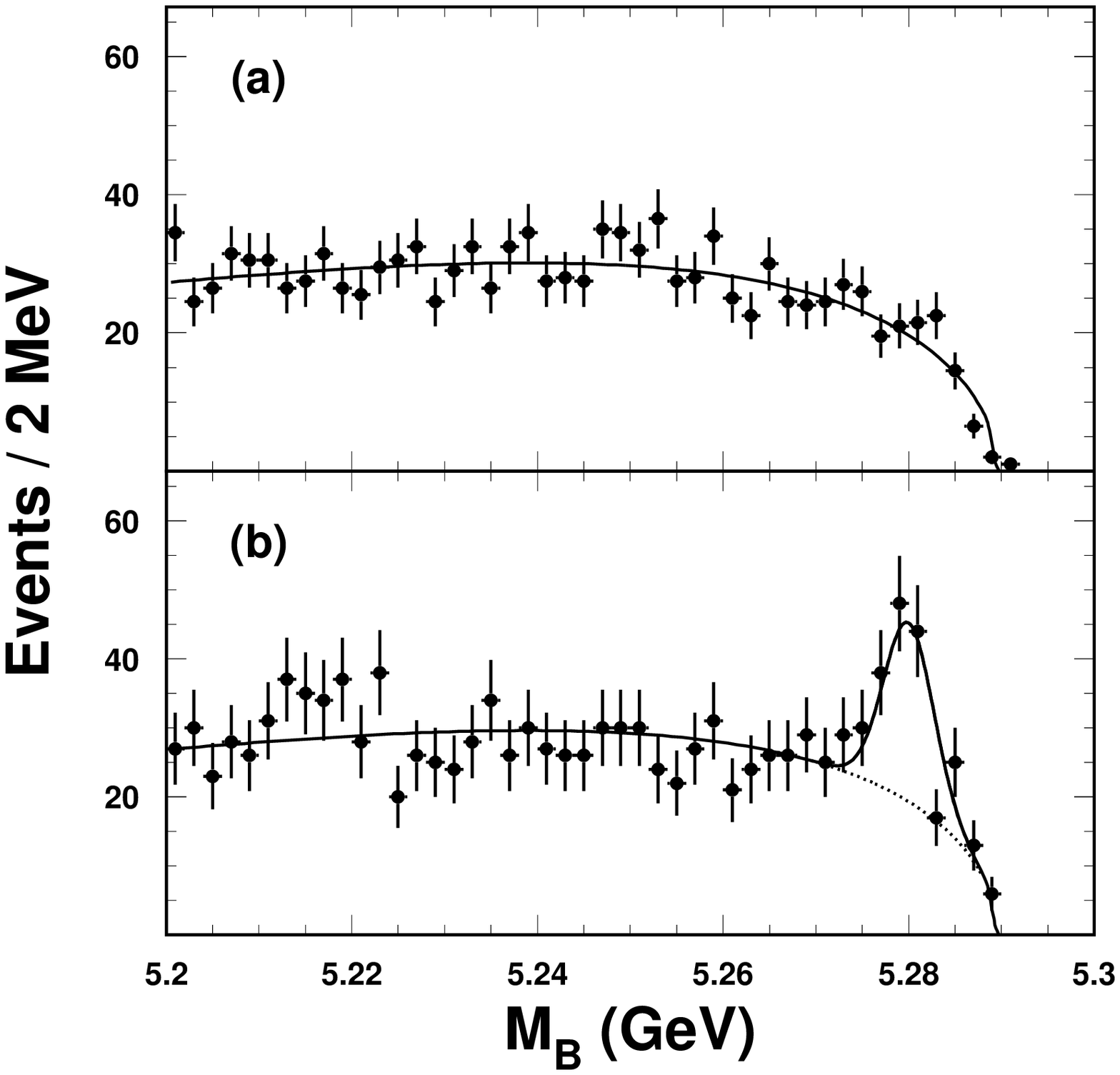,height=4in}} 
\vspace{-1.0cm} 
\caption{ \label{bm_omega_D0}The $B$ candidate mass spectra for 
the final state 
$D^{0}\omega\pi^-$, with $D^o\to K^-\pi^+$. (a) for $\Delta E$ 
sidebands, and (b) for $\Delta E$ consistent with zero. The 
vertical scale
in (a) was multiplied by 0.5 to facilitate comparison.  The curve 
in (a) is a 
fit to the background distribution described in the text, while in 
(b) the 
shape from (a) is used with the normalization allowed to float and 
a signal 
Gaussian of width 2.7 MeV is added. 
} 
\end{figure}

We start with the $D^o\to K^-\pi^+$ decay mode, for events in the 
$\omega$ peak. 
We show the candidate $B$ mass distribution, $M_B$, for
$\Delta E$  in the side-bands from -7.0 
to -3.0$\sigma$ and 7.0 to 3.0$\sigma$ 
on Fig.~\ref{bm_omega_D0}(a). The $\Delta E$ resolution 
is 18 MeV ($\sigma).$ This gives a good representation of the 
background in the signal region.  
We fit this distribution with a shape given as  
\begin{equation}\label{eq:background} 
back(r)=p_1 r\sqrt{1-r^2}e^{-p_2(1-r^2)}~~~, 
\end{equation} 
where $r=M_B/5.2895$, and the $p_i$ are parameters given by the 
fit.

 We next view the $M_B$ distribution for events having  $\Delta E$  
within 2$\sigma$ around zero in 
Fig.~\ref{bm_omega_D0}(b). This distribution is fit with a 
Gaussian signal  
function of width 2.7 MeV and the background function found above 
whose  
normalization is allowed to vary. We find 88$\pm$14 events in the 
signal peak.  
 
\begin{figure}[htb] 
\centerline{\epsfig{figure=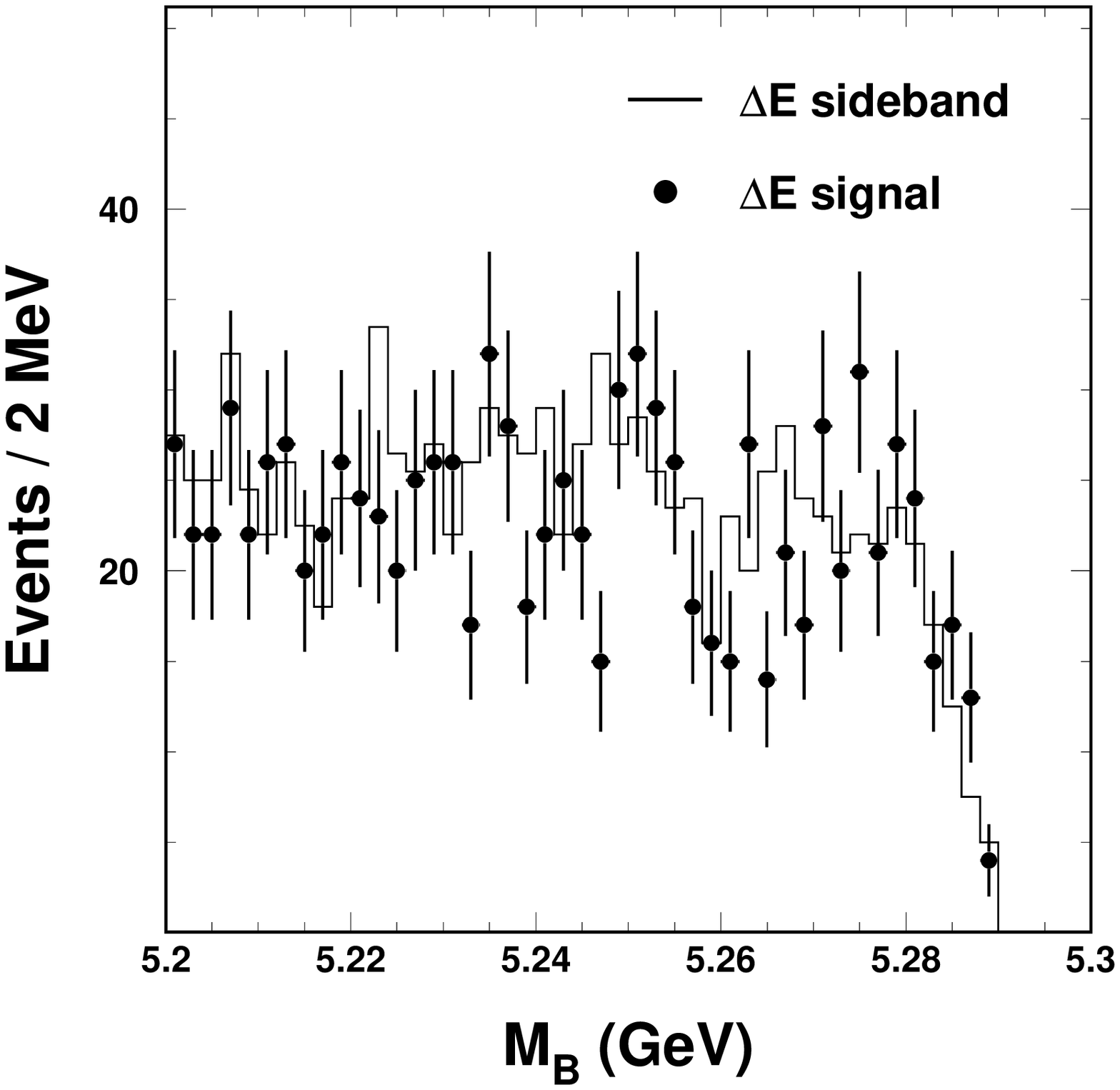,height=4in}} 
\vspace{-1.0cm} 
\caption{ \label{bm_not_omega_D0}The $B$ candidate mass spectra 
for the final state 
$D^{0}\omega\pi^-$, with $D^o\to K^-\pi^+$ and $\omega$ sidebands
(a) for $\Delta E$ sidebands and (b) for $\Delta E$ consistent 
with zero.
} 
\end{figure}

We repeat this procedure for events in the $\omega$ sidebands. We 
use for
our $\chi^2_b$ definition pseudo-$\omega$ masses in the sideband
intervals.
We show the $M_B$ distribution for events in the $\Delta E$ 
sideband,
defined above, and those having  $\Delta E$  within 2$\sigma$ 
around zero in 
Fig.~\ref{bm_not_omega_D0}. We find no significant signal.

 \subsection{$\overline{B}^o\to D^+\omega\pi^-$ Signal} 
 
\begin{figure}[htb] 
\centerline{\epsfig{figure=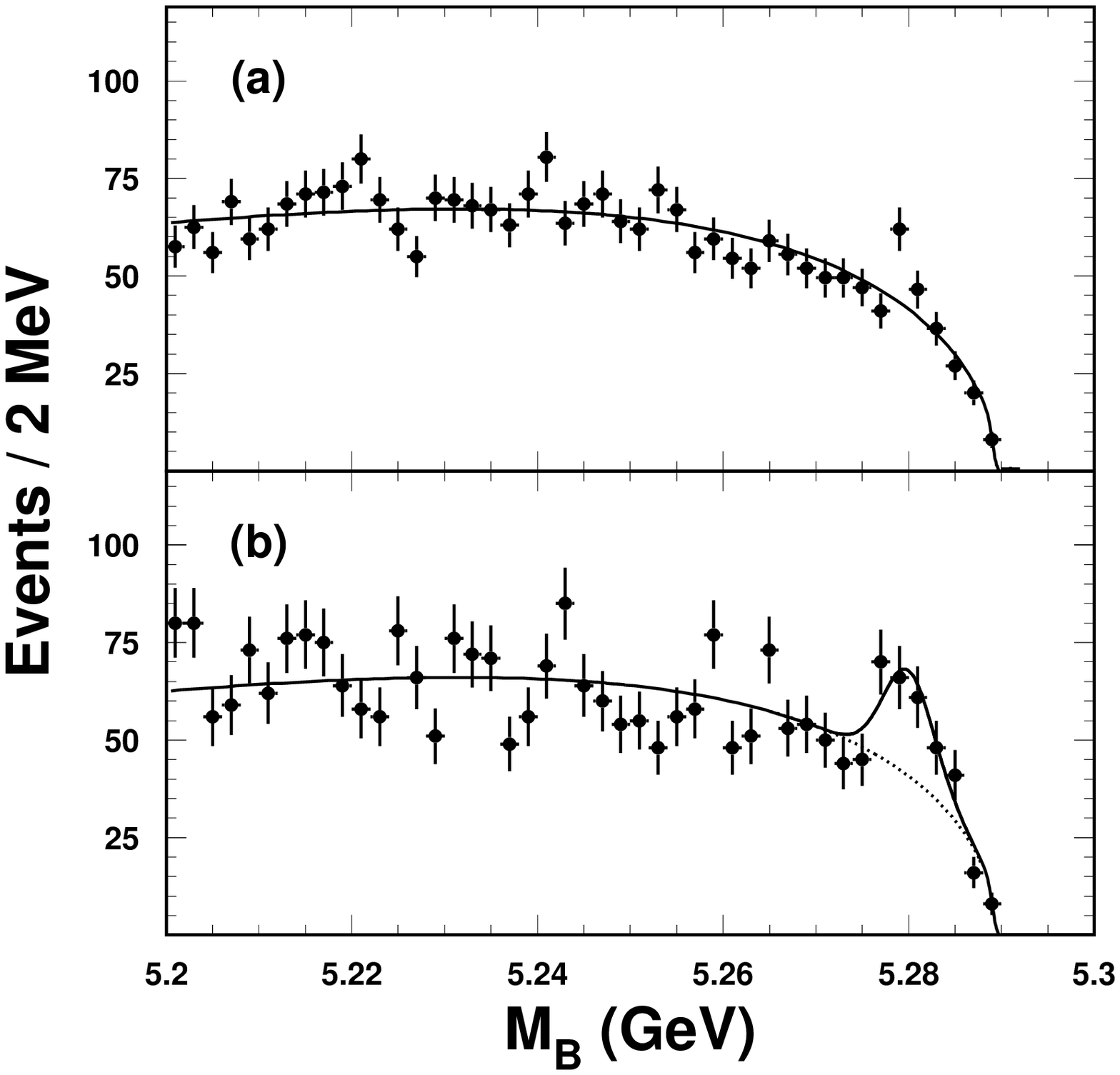,height=4in}} 
\vspace{-1.0cm} 
\caption{ \label{bm_omega_D+}The $B$ candidate mass spectra for 
the final state 
$D^{+}\omega\pi^-$, with $D^+\to K^-\pi^+\pi^+$ (a) for $\Delta E$ 
sidebands and (b) for $\Delta E$ consistent with zero. The 
vertical scale
in (a) was multiplied by 0.5 to facilitate comparison. The curve 
in (a) is a 
fit to the background distribution described in the text, while in 
(b) the 
shape from (a) is used with the normalization allowed to float and 
a signal 
Gaussian of width 2.7 MeV is added. 
} 
\end{figure}

The same procedure followed for the $D^o$ final state is used for the $D^+$
final state.  
We show the candidate $B$ mass distribution, $M_B$, for events in the  
$\Delta E$ side-band
 on Fig.~\ref{bm_omega_D+}(a). The $\Delta E$ resolution 
is 18 MeV ($\sigma).$ This gives a good representation of the 
background in the signal region.  
We fit this distribution with a shape given in equation~\ref{eq:background}. 

 We next view the $M_B$ distribution for events having  $\Delta E$  
within 2$\sigma$ around zero in Fig.~\ref{bm_omega_D+}(b). 
This distribution is fit with a Gaussian signal  
function of width 2.7 MeV and the background function found above 
whose  
normalization is allowed to vary. We find 91$\pm$18 events in the 
signal peak.  
 
\begin{figure}[htb] 
\centerline{\epsfig{figure=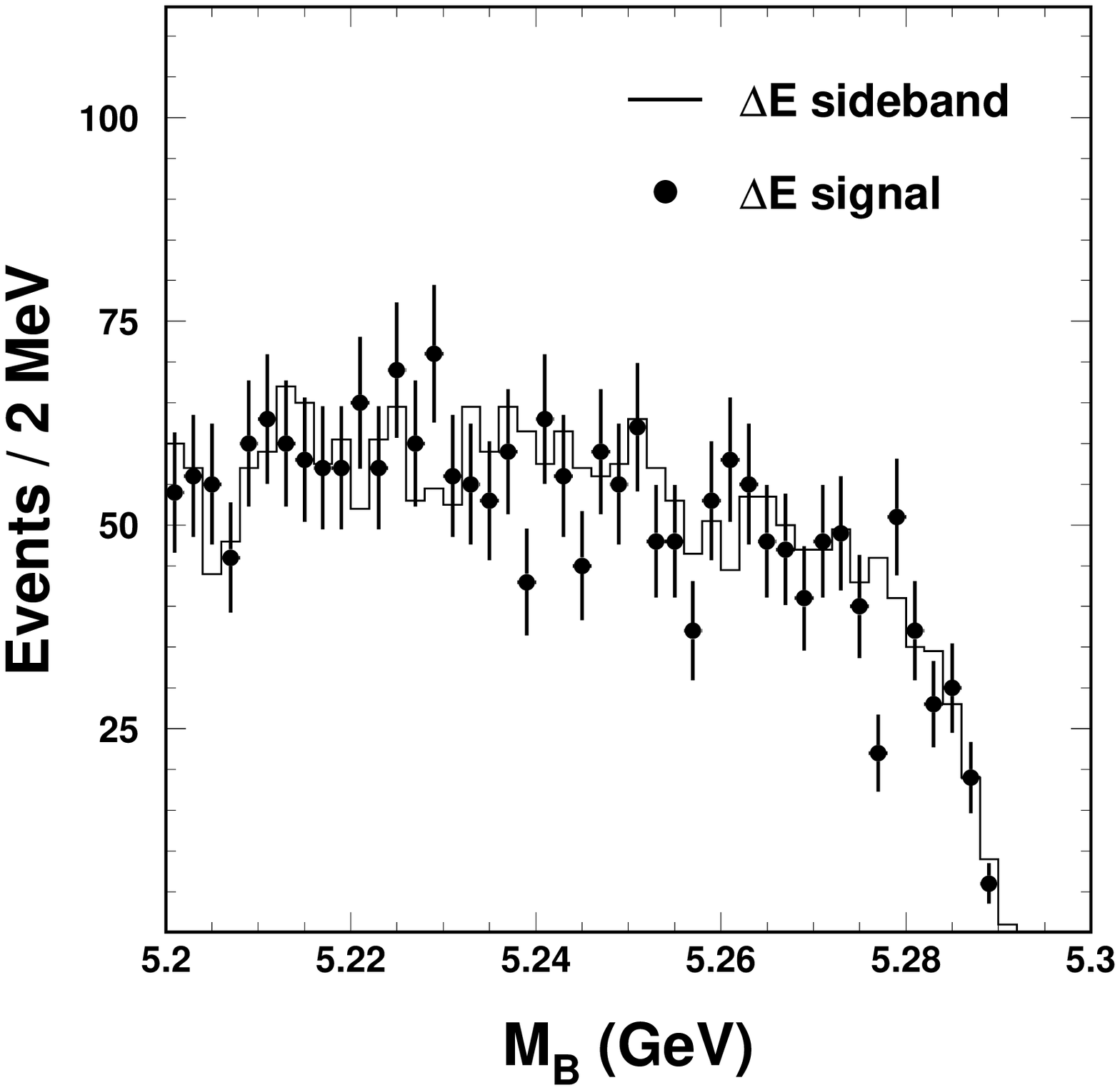,height=4in}} 
\vspace{-1.0cm} 
\caption{ \label{bm_not_omega_D+}The $B$ candidate mass spectra 
for the final state 
$D^{+}\omega\pi^-$, with $D^+\to K^-\pi^+\pi^+$ and $\omega$ 
sidebands
(a) for $\Delta E$ sidebands and (b) for $\Delta E$ consistent 
with zero.
} 
\end{figure}

We repeat this procedure for events in the $\omega$ sidebands. 
We show the $M_B$ distribution for both $\Delta E$ sidebands and  
$\Delta E$  
within 2$\sigma$ around zero in Fig.~\ref{bm_not_omega_D+}.

 There is no evidence of any signal in the $\omega$ sideband plot, 
 leading to the conclusion that the signal is associated purely 
with $\omega$.

\subsection{Branching Fractions}

We determine the branching ratios, shown in 
Table~\ref{table:Domgpievents},
by performing a Monte Carlo simulation of the
efficiencies in the two modes.
 We use the current particle data group 
values for the relevant $\omega$, $D^{+}$ and $D^o$ branching 
ratios of 
(88.8$\pm$0.7)\% ($\omega\to \pi^+\pi^-\pi^o$), 
(9.0$\pm$0.6)\% ($D^{+}\to K^-\pi^+\pi^+$) and  
(3.85$\pm$0.09)\% ($D^o\to K^-\pi^+$) \cite{PDG}.
 The efficiencies listed in the table do not include these 
branching ratios \cite{spineff}.

\begin{table}[hbt]  
\begin{center}  
\caption{Branching Fractions for the $D\omega\pi^-$ final state}  
\label{table:Domgpievents}  
\begin{tabular}{lccc}\hline\hline  
$D^o$ Decay Mode & Fitted \# of events& Efficiency &Branching 
Fraction (\%)\\\hline  
$K^-\pi^+$&               88$\pm$14   & 0.064 &  
0.41$\pm$0.07$\pm$0.04   \\  
$K^-\pi^+\pi^+$ &         91$\pm$18   & 0.046 &  
0.28$\pm$0.05$\pm$0.03   \\\hline  
\end{tabular}  
\end{center}  
\end{table}

The systematic error arises mainly from our lack of knowledge 
about the  
tracking and $\pi^o$ efficiencies. We assign errors of $\pm$2.2\% 
on the  
efficiency of each charged track,  and  
$\pm$5.4\% for the $\pi^o$. The error due to the 
background shape is evaluated  
in three ways. First of all, we change the background shape by 
varying the 
fitted parameters by 1$\sigma$. This results in a change of 
$\pm$5.0\%. Secondly, 
we allow the shape, $p_2$, to vary (the normalization, $p_1$, was 
already 
allowed to vary). This results in 5.5\% increase in the number of 
events. 
Finally, we choose a different background function 
\begin{equation} 
back'(r)=p_1 r\sqrt{1-r^2}\left(1+p_2 r + p_3 r^2 +p_4 r^3 
\right)~~~, 
\end{equation} 
and repeat the fitting procedure. This results in a 1.0\% decrease 
in the 
number of events. Taking a conservative estimate of the systematic 
error due to 
the background shape we arrive at $\pm$5.5\%.

\subsection {The $\omega\pi^-$ System} 
 
For all subsequent discussions we add the $D^o$ and $D^+$ final 
states together. 
We select sample of $\omega$'s in the $\pi^+\pi^-\pi^o$ mass 
window of 782$\pm$20 MeV using only combinations having $r<0.7$ 
in the Dalitz plot. 
 
\begin{figure}[htb] 
\centerline{ \epsfig{figure=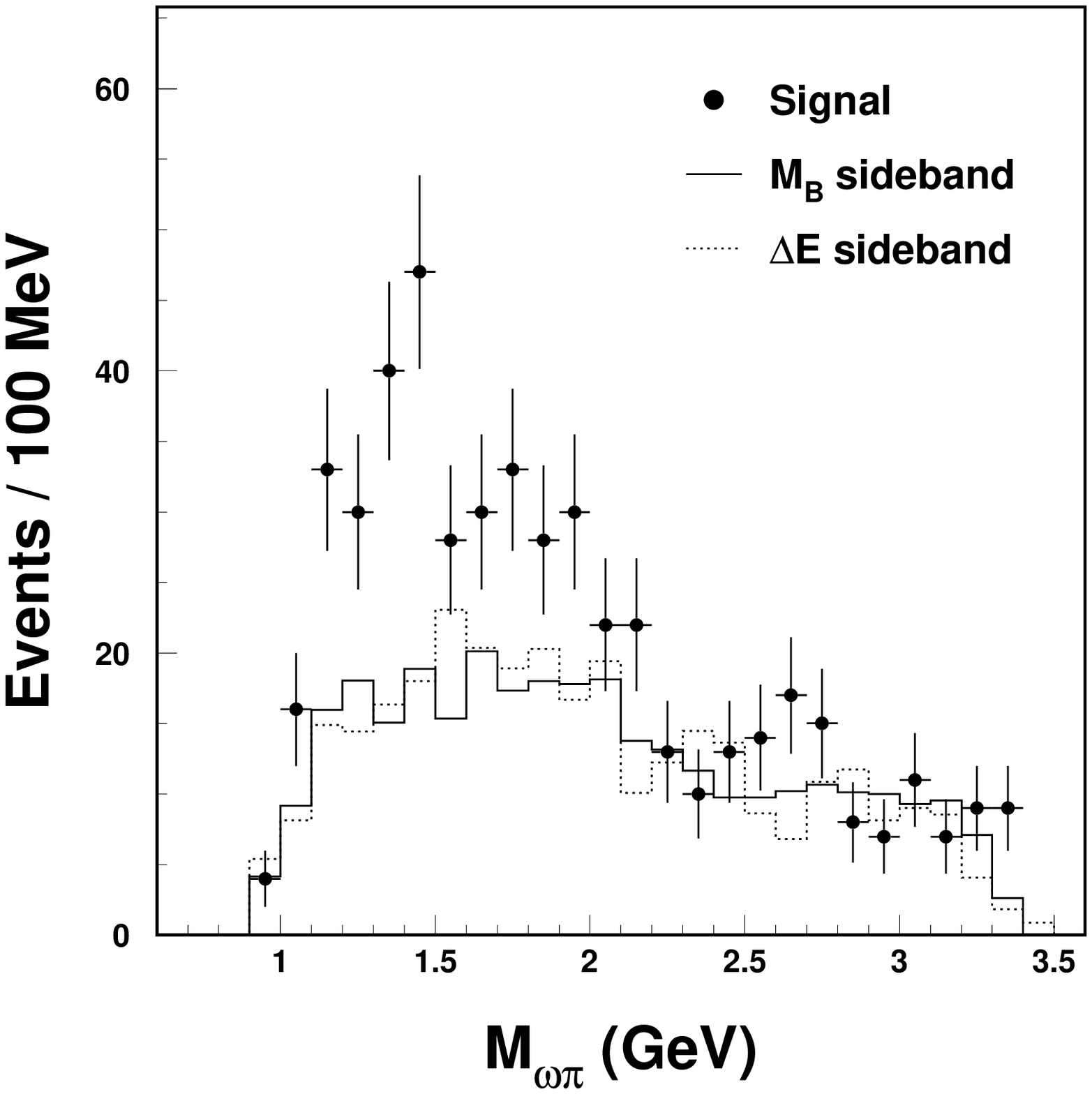,height=3.7in}
             \epsfig{figure=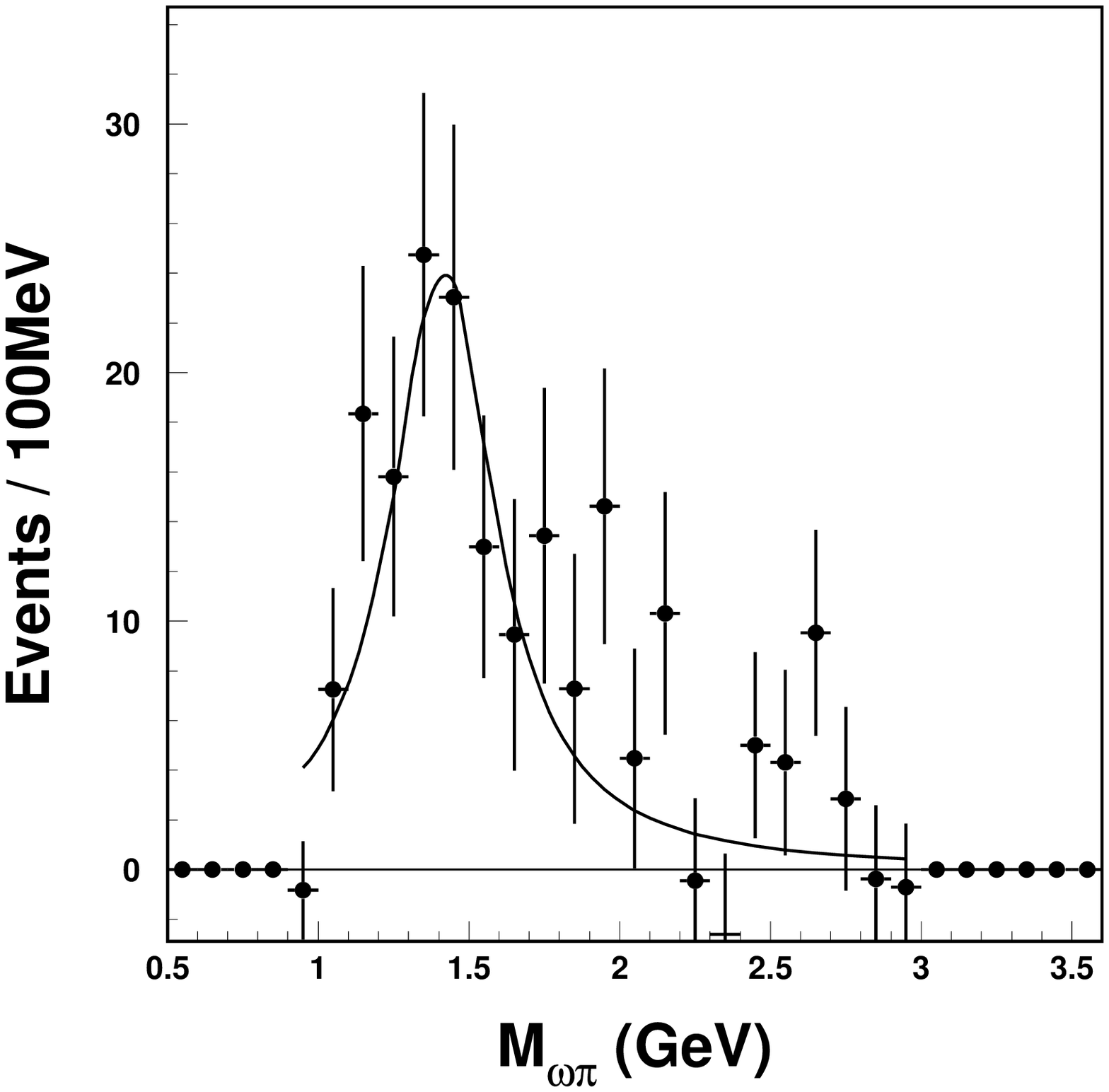,height=3.7in} }
\vspace{-1.0cm}
\caption{ \label{m_omega_pi} The invariant mass spectra of 
${\omega\pi^-}$ for the
final state $D\omega\pi^-$ for both ${D}$ decay modes. (left) The 
solid histogram
is the background estimate from the $M_B$ lower sideband and the 
dashed histogram
is from the $\Delta E$ sidebands; both are normalized to the 
fitted number of
background events. (right) The mass spectrum determined from 
fitting the $M_B$ 
distribution and fit to a Breit-Wigner function.}
\end{figure}

In Fig.~\ref{m_omega_pi} we show the $\omega\pi^-$ mass spectrum 
in the left-side plot. 
The solid histogram shows events from the  
lower $M_B$ sideband region (5.203 - 5.257 GeV) suitably 
normalized. The dotted histogram shows  
the background estimate from the $\Delta E$ sidebands, again 
normalized.  
In the signal distribution there is a wide structure around   
1.4 GeV, that is inconsistent with background. We re-determine the 
$\omega\pi^-$ 
mass distribution by fitting the $M_B$ distribution in bins of 
$\omega\pi^-$ mass, and this is shown on the right-side.  Fitting 
to a Breit-Wigner function, 
we find a peak value of 1415$\pm$43 MeV and a width of 419$\pm$110 
MeV. It should be kept in mind that this particular signal 
shape is assumed to be correct. Other resonant or non-resonant 
contributions could affect the mass and width.

This structure appears identical to the one we observed in
$\overline{B}\to D^{*}\omega\pi^-$ decays.

\subsection{Angular Distributions in $D\omega\pi^-$}
\label{sec:Dangular}
We can determine the spin and parity of the $A^-$ 
particle by
studying the angular distributions characterizing its decay 
products. 
The decay chain that we are considering is $B\rightarrow A\ D$; 
$A\rightarrow
\omega \pi$ and $\omega \rightarrow \pi ^+ \pi ^- \pi ^o$. The 
helicity formalism \cite{helicity} is 
generally used in the analysis of these 
sequential decays. This formalism is well suited to relativistic
problems involving particles with spin $\vec{S}$ and momentum 
$\vec{p}$ because the helicity operator $h=\vec{S}\cdot \vec{p}$ 
is invariant under both rotations and boosts along $\hat{p}$. 

\begin{figure}
\centerline {\epsfxsize=3.5in  
\epsffile{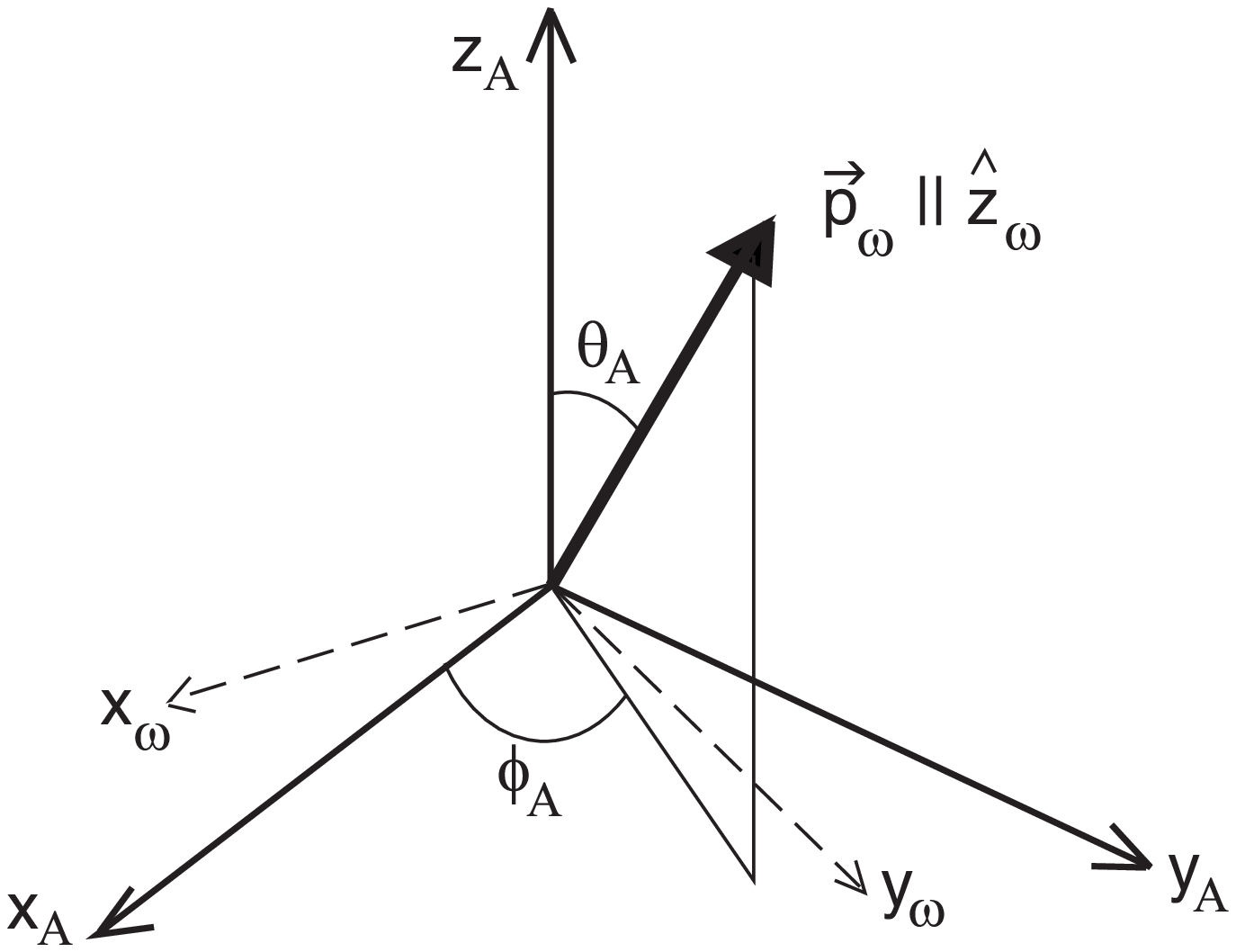}}
\caption{Relationship between the $A$ rest frame $x_Ay_Az_A$ and 
the $\omega$
rest frame $x_{\omega}y_{\omega}z_{\omega}$. $x_A$ and 
$x_{\omega}$ lie in 
the same plane.}
\label{eul-angles}
\end{figure} 

There are two relevant reference frames. The first one, that we 
will 
define $x_Ay_Az_A$ is the rest frame of the $A$ particle, with the
$\hat{z_A}$ axis 
pointing in the $A$ direction of motion in the $B$ rest frame. The 
second
one, $x_{\omega}y_{\omega}z_{\omega}$, is
related to $x_Ay_Az_A$ by the rotation through 3 Euler angles 
$\phi _A,\theta _A,-\phi _A$,
as shown in Fig.~\ref{eul-angles}. The angle $\phi _A$ defines the 
orientation 
of the plane containing the $\omega$ direction in the A rest frame 
and the 
$\hat{z_A}$ axis with respect to the 
$\hat{x_A}-\hat{z_A}$ 
plane. The $\hat{x_A}$ direction is arbitrary. The angles 
$\theta _{\omega}$ and $\phi _{\omega}$ define the orientation of 
the
$\omega$ decay plane in the $\omega$ rest frame. Note that the 
$A$ decay 
plane has an azimuthal angle $\phi _A$ both in the $x_Ay_Az_A$ and 
in 
the $x_{\omega}y_{\omega}z_{\omega}$ references. As the angle 
$\phi _A$ is 
arbitrary, the only angle that has a physical meaning is $\chi = 
\phi _A - \phi _{\omega}$, the opening angle between the $A$ decay 
plane and
 the $\omega$ decay plane.

Both the $B$ meson and the $D$ meson are pseudoscalar, therefore 
their helicity is 0. Thus $A$ will be longitudinally polarized 
independently of its spin. In order to calculate the decay 
amplitude for this
process, we need to sum over the $\omega$ helicity states:

\begin{equation}
{\cal A} = \Sigma _{\lambda _{\omega}} D^{\star J_A}_{0\lambda
_{\omega}}(\phi _A,\theta _A, -\phi _A)
 D^{\star 1}_{\lambda_{\omega}0}(\phi _{\omega},\theta_{\omega}, -
\phi _
{\omega}) B_{\lambda_{\omega}0},
\end{equation}
here $D^{\star 1}_{\lambda_{\omega}}(\phi _A,\theta _A,-\phi _A)$ 
is the 
rotation 
matrix that relates the $x_Ay_Az_A$  and the 
$x_{\omega}y_{\omega}z_{\omega}$ frames
and $ D^{\star 1}_{\lambda_{\omega}}(\phi 
_{\omega},\theta_{\omega}, -\phi
_{\omega})$ is the rotation matrix relating the 
$x_{\omega}y_{\omega}z_{\omega}$ and the direction of the normal 
to the
$\omega$ decay plane $\hat{n}(\theta _{\omega},\phi_{\omega})$.

In general, there are three helicity amplitudes that contribute to 
this decay:
$B_{10}$ and 
$B_{-10}$, corresponding to a transverse 
$\omega$ polarization, and $B_{00}$, corresponding to a 
longitudinal $\omega$
polarization. This expression can be simplified by observing that 
 $A\rightarrow \omega \pi$ is a strong decay and thus conserves
parity. Thus, the helicity amplitudes are related by the equation:

\begin{equation}
B_{10}=(-1)^{1-S(A)}\eta_{A}\eta_{\omega}\eta_{\pi}B_{-10}, 
\label{tran}
\end{equation}
\begin{equation}
B_{00}=(-1)^{1-S(A)}\eta_{A}\eta_{\omega}\eta_{\pi}B_{00}.
\label{long}
\end{equation}
where $S(A)$ is the spin of particle $A$ and $\eta 
_A,\eta_{\omega}$ and
$\eta_{\pi}$ represent the intrinsic parity of the decaying 
particle and its
decay products, respectively.

Eq.~\ref{tran} relates the two transverse helicity
amplitudes, while Eq.~\ref{long} forbids the presence of a 
longitudinal
component under certain conditions. For example, if $A$ is a $1^-$ 
object,
$\omega$ has transverse polarization and $B_{-10}= - B_{10}$. When 
the sign
in Eqs.~\ref{tran}-\ref{long} is positive, two parameters 
determined by
the hadronic matrix element affect the angular distribution and 
thus we 
cannot fully determine it only on the basis of our assumptions on 
the 
$A$ spin parity. We have carried out the calculation of the 
predicted angular
distributions including spin assignment for $A$ up to 2. The 
predicted
angular distributions are summarized in Table \ref{diff-ang}.

\begin{table}
\caption{Differential angular distributions (modulo a 
proportionality
constant) predicted for different spin assignments.}
\label{diff-ang}
\begin{center}
\begin{tabular}{lc}
\hline
\hline
$J^{\eta}$ & ${d\sigma}/{d\cos{\theta _A}d\cos{\theta 
_{\omega}}d\chi}$
\\
\hline
$0^-$ &   $|B_{00}|^2 \cos^2{\theta_{\omega}}$  \\
$1^-$ &   $|B_{10}|^2 \sin^2{\theta _A}\sin^2{\theta 
_{\omega}}\sin^2{\chi}$ \\
$1^+$ &   $|B_{10}|^2 \sin^2{\theta _A}\sin^2{\theta 
_{\omega}}^2\cos{\chi}^2 +
           |B_{00}|^2 \cos^2{\theta _A}\cos^2{\theta_{\omega}}$ \\
  &          $ - 1/2Re(B_{10}B_{00}^{*})\sin{2\theta _A}
             \sin{2\theta _{\omega}}\cos{\chi}$ \\
$2^-$ & $3|B_{10}|^2\sin^2{2\theta 
_A}\sin^2{\theta_{\omega}}\cos^2{\chi}
        +|B_{00}|^2(3\cos^2{\theta _A}-1)^2\cos^2{\theta 
_{\omega}}$ \\
      & $ -\sqrt{3}Re(B_{10}B_{00}^{*})\sin{2\theta 
_A}(3\cos^2{\theta _A}-1)
	\sin{2\theta _{\omega}}\cos{\chi}$\\
$2^+$ &    $3/4 |B_{10}|^2 \sin^2{2\theta _A}\sin^2{\theta 
_{\omega}}\sin^2{\chi}$ 
\\ \hline
\end{tabular}
\end{center}
\normalsize
\end{table}

The statistical accuracy of our data sample is not sufficient to 
do a 
simultaneous fit of the joint angular distributions shown above. 
Thus only
the projections along the $\theta _A$, $\theta _{\omega}$ and 
$\chi$ are fitted, integrating out the remaining degrees of 
freedom. Table~\ref{proj1},
gives the analytical form for these projections.

\begin{table}
\caption{Projection of the angular distributions along the 
$\cos{\theta _A}$, $\cos{\theta _{\omega}}$ and $\chi$ axes.}
\label{proj1}
\begin{center}
\begin{tabular}{lccc}
\hline
\hline
$J^{\eta}$ & $d\sigma/d\cos{\theta _A}$ & 
$d\sigma/d\cos{\theta_{\omega}}$
& $d\sigma/d{\chi}$ \\
\hline
$0^-$ & $\frac{4\pi}{3}|B_{00}|^2$  & $4\pi 
|B_{00}|^2\cos^2{\theta _{\omega}}$ & 
$4/3|B_{00}|^2$ \\
$1^-$ & $\frac{4\pi}{3} |B_{10}|^2\sin^2{\theta _A}$&
$\frac{4\pi}{3} |B_{10}|^2\sin^2{\theta_{\omega}}$ &  
$\frac{8}{9}|B_{10}|^2\sin^2{\chi}$\\
$1^+$ & $\frac{4\pi}{3} (|B_{10}|^2 \sin^2{\theta _A}$ 
& $\frac{4\pi}{3} (|B_{10}|^2\sin^2{\theta _{\omega}}$ &  
$ \frac{4}{9}(4|B_{10}|^2\cos^2{\chi}$   \\
  & $+|B_{00}|^2\cos^2{\theta _A})$   &  $+|B_{00}|^2\cos^2{\theta 
_{\omega}})$   &  $+|B_{00}|^2)$   \\
$2^-$ &
$\frac{4\pi}{3}(3|B_{10}|^2\sin^2{2\theta_A}$  
& $\frac{16\pi}{5}(|B_{10}|^2\sin^2{\theta _{\omega}}$
& $4|B_{10}|^2\cos^2{\chi}$    \\
  & $+|B_{00}|^2(3\cos^2{\theta_A}-1)^2)$   & 
$+|B_{00}|^2\cos^2{\theta _{\omega}}) $    &  $+|B_{00}|^2$   \\
$2^+$ & $\pi |B_{10}|^2\sin^2{2\theta _A}$ &$\frac{4\pi}{5} 
|B_{10}|^2
\sin^2{\theta _{\omega}}$  & 
$\frac{16}{15}|B_{10}|^2\sin^2{\chi}$ \\
\hline
\end{tabular}
\end{center}
\normalsize
\end{table}

\begin{figure}[htb] 
\centerline{\epsfig{figure=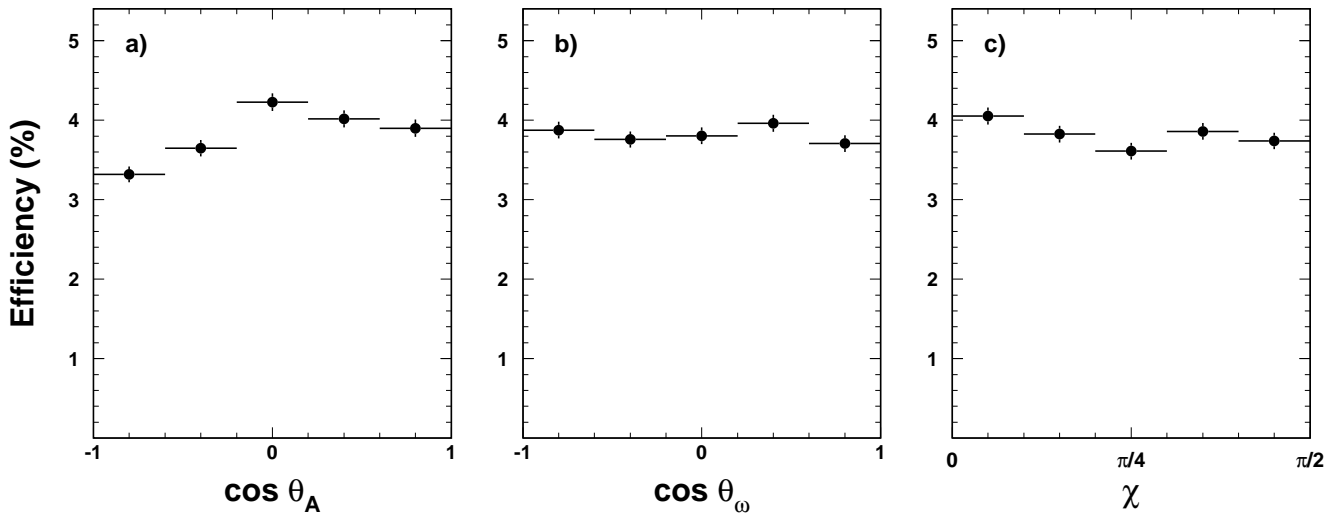,height=3.0in}} 
\vspace{-1mm} 
\caption{ \label{efficiency} Reconstruction efficiency dependence 
on (a) $\cos\theta_A$, 
(b) $\cos\theta_\omega$, and (c) $\chi$.}
\end{figure}

  
We determine the projections of these angular distributions by 
fitting  
the $M_B$ distribution as a function of the various angular 
quantities
$\cos\theta_A$, $\cos\theta_{\omega}$, $\chi$. We restrict the 
$\omega\pi^-$
mass range to be between 1.1 and 1.7 GeV, containing 104 events. 
In order to fit the angular distribution with theoretical 
expectations,
we must the correct the data for acceptances. 
We determine the acceptance correction by comparing the Monte 
Carlo generated
angular distributions with the reconstructed distributions. 
The angular dependent efficiencies are shown in 
Fig.~\ref{efficiency}.
 
\begin{figure}[htbp] 
\centerline{\epsfig{figure=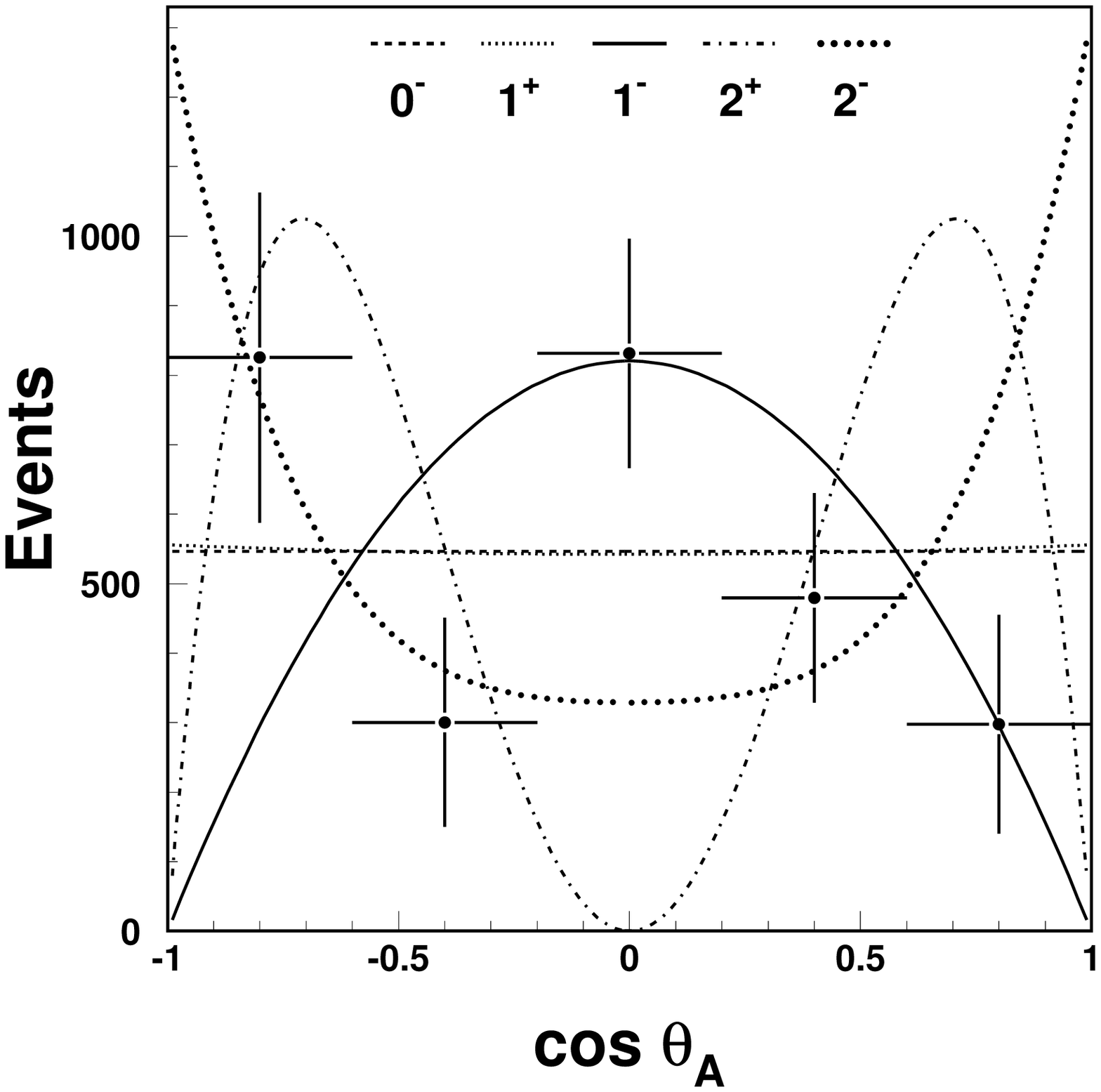,height= 3.6in}\hspace{-4mm} 
\epsfig{figure=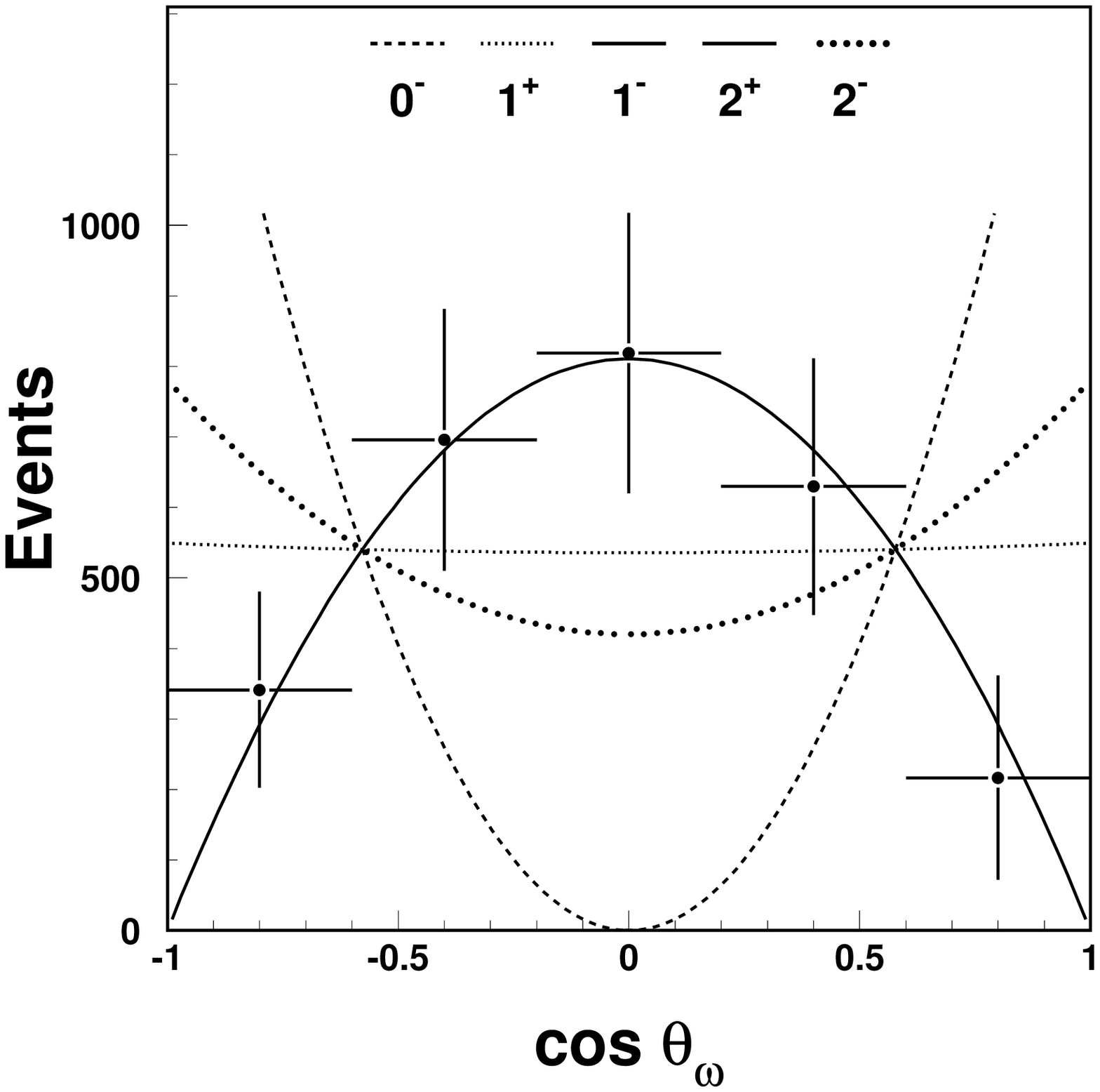,height=3.6in}}
\centerline{\epsfig{figure=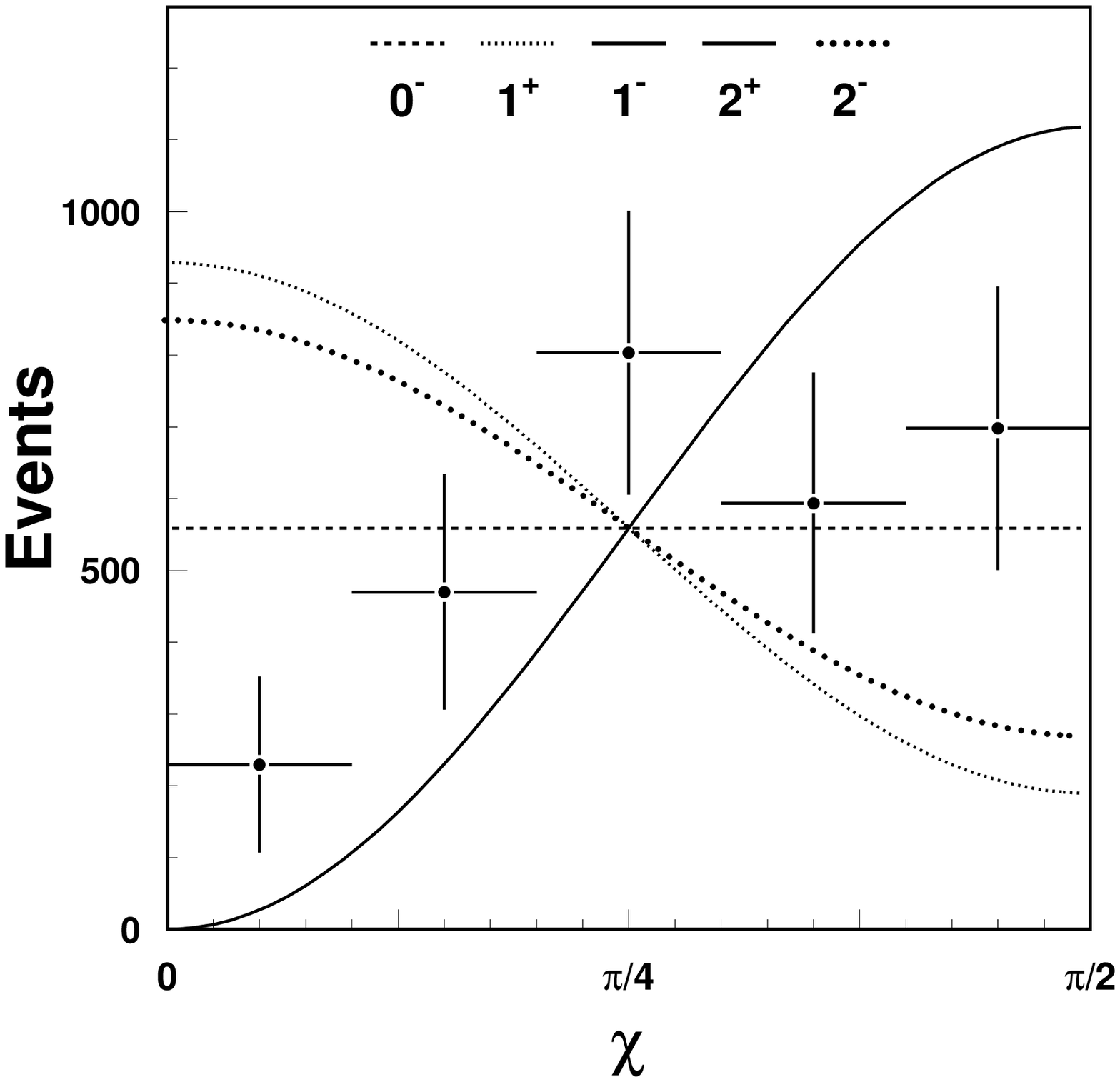,height=3.6in}}
\caption{ \label{angle_cosa} The angular distribution of 
$\theta_A$ (top-left),
$\theta_\omega$ (top-right) and $\chi$ (bottom).
     The curves show the best fits to the data for 
      for different $J^P$ assignments. (The $0^-$ and $1^+$ are 
almost
      indistinguishable in $\cos\theta_A$, while the $1^-$ and 
$2^+$ are
      indistinguishable in $\cos\theta_{\omega}$ and $\chi$.}
\end{figure}

The corrected angular distributions are shown in 
Fig.~\ref{angle_cosa}.
The data are fit to the expectations for the various $J^P$ 
assignments. 
For the $0^-$, $1^-$ and $2^+$ assignments, the curves have a 
fixed shape. 
For the $1^+$ and $2^-$ assignments we let the ratio between the 
longitudinal
 and transverse amplitudes vary to best fit the data. 
We notice that the $\omega$ polarization is very clearly 
transverse 
($\sin^2\theta_{\omega}$) and that infers a $1^-$ or $2^+$ 
assignment.

We list in Table~\ref{table:chi2_of_fit} the $\chi^2/dof$ for the 
different 
$J^P$ assignments. 
The $1^-$ assignment is preferred, having a  $\chi^2/dof$ of 1.7. 
The other 
assignments are clearly ruled out. 
The probability that we have a correct solution and $\chi^2/dof$ 
is 1.7 or
greater is 3.8\% \cite{systematics}.

\begin{table}[hbt]  
\begin{center}  
\caption{The $\chi^2$ of angular fits}
\label{table:chi2_of_fit}  
\begin{tabular}{cccccc}\hline\hline  
              &  $0^-$ &  $1^+$ &  $1^-$ &  $2^+$ & $2^-$ \\\hline  
$\chi^2/dof$ &  7.0 & 4.5 & 1.7 & 3.2 & 5.3 \\
probability   &  $1.9\times 10^{-15}$   &  $3.3\times 10^{-8}$ & 
3.8\% 
              &  $2.7\times 10^{-5}$    & $3.3\times 10^{-10}$ \\
\hline  
\end{tabular}  
\end{center}  
\end{table}

\section{Discussion of Nature of the $A^-$}

We have found a $1^-$ object decaying into $\omega\pi^-$.
A simple Breit-Wigner fit assuming a single resonance and no 
background gives a mass of 1418$\pm$26 MeV with an intrinsic width 
of 382$\pm$41 MeV. 
(The individual measurements are listed in 
Table~\ref{table:maswd}). We have evaluated the systematic errors 
due to changes in the parameterization of the background shape in 
when fitting the $M_B$ distributions in bins of $\omega\pi^-$ mass. 
This leads to an error in the mass of 19 MeV and 32 MeV in the width. 

\begin{table}[hbt]
\begin{center}
\caption{Measured $\rho'$ Mass and Width}
\label{table:maswd}
\begin{tabular}{lrr}\hline\hline
Mode & Mass (MeV) & Width (MeV)\\\hline
$\overline{B}^o\to D^{*+}\omega\pi^-$ & 1432$\pm$37& 376$\pm$47  
\\
${B}^-\to D^{*o}\omega\pi^-$  & 1367$\pm$75 & 439$\pm$135     
\\
${B}\to D\omega\pi^-$ & 1415$\pm$43& 419$\pm$110 \\\hline
Average  &1418$\pm$26   & 388$\pm$41 \\
\hline
\end{tabular}
\end{center}
\end{table}
Other possible changes in the mass and width are more difficult to evaluate. 
They include different 
Breit-Wigner parameterizations and the possibility of other resonant or 
non-resonant components in the $\omega\pi^-$ mass distributions.

Signals for $\omega\pi^-$
resonances have been detected before below 1500 MeV.
There is a  well established axial-vector state the b$_1$(1235) 
with mass 1230 MeV and 
width 142 MeV. Data on vector states, excited $\rho$'s, is 
inconsistent. Clegg and 
Donnachie \cite{Clegg}  have reviewed $\tau^-\to(4\pi)^-
\bar{\nu}$, 
$e^+e^-\to\pi^+\pi^-$ and $e^+e^-\to\pi^+\pi^+\pi^-\pi^-$ data, 
including the 
$\omega\pi$ final state. Their best explanation is that of two 
$1^-$ states at
1463$\pm$25 MeV and 1730$\pm$30 MeV with widths 311$\pm$62 and 
400$\pm$100 MeV,
respectively. Only the lighter one decays into $\omega\pi$. The 
situation is quite complex, however. They conclude that 
these states must be mixed with non-$q\overline{q}$ states in 
order to explain their decays widths. There is also an observation 
of a wide, 300 MeV, 
$\omega\pi^o$ state in photoproduction at 1250 MeV \cite{Aston}, 
that is dominantly the $b_1$(1235) \cite{Brau} with possibly some
$1^-$ in addition. 
Our state is consistent with the lower mass $\rho'$. We do not
seem to be seeing significant production of the higher mass state 
into $\omega\pi^-$, as expected.

Several models predict the mass and decay widths of excited $\rho$ 
and 
$\omega$ mesons . For example, according to Godfrey 
and Isgur \cite{GImodel} 
the first radial excitation of the $\rho$ is at 1450 MeV. There is 
a large 
variation among the models, however, on prediction of the relative 
decays 
widths ranging from no $\pi\pi$ to $\pi\pi$ being equal to 
$\omega\pi$ \cite{Othermodels}.

Since we have observed a wide $1^-$ state in the mass region where 
the $\rho'$ is expected, the most natural explanation is that we 
are observing the $\rho'$ for the first time in $B$ decays.

We note that $\tau^-$ lepton decays into $\omega\pi^-$ have been observed,
and the $1^-$ spin-parity definitely established \cite{CLEOtau}. 
However, the relatively low mass of the $\tau^-$ distorts the 
the mass spectrum significantly, and makes it difficult to extract the
$\rho'$ mass and width \cite{tautopipi}. 

\begin{figure}[tbh]
\centerline{\epsfig{figure=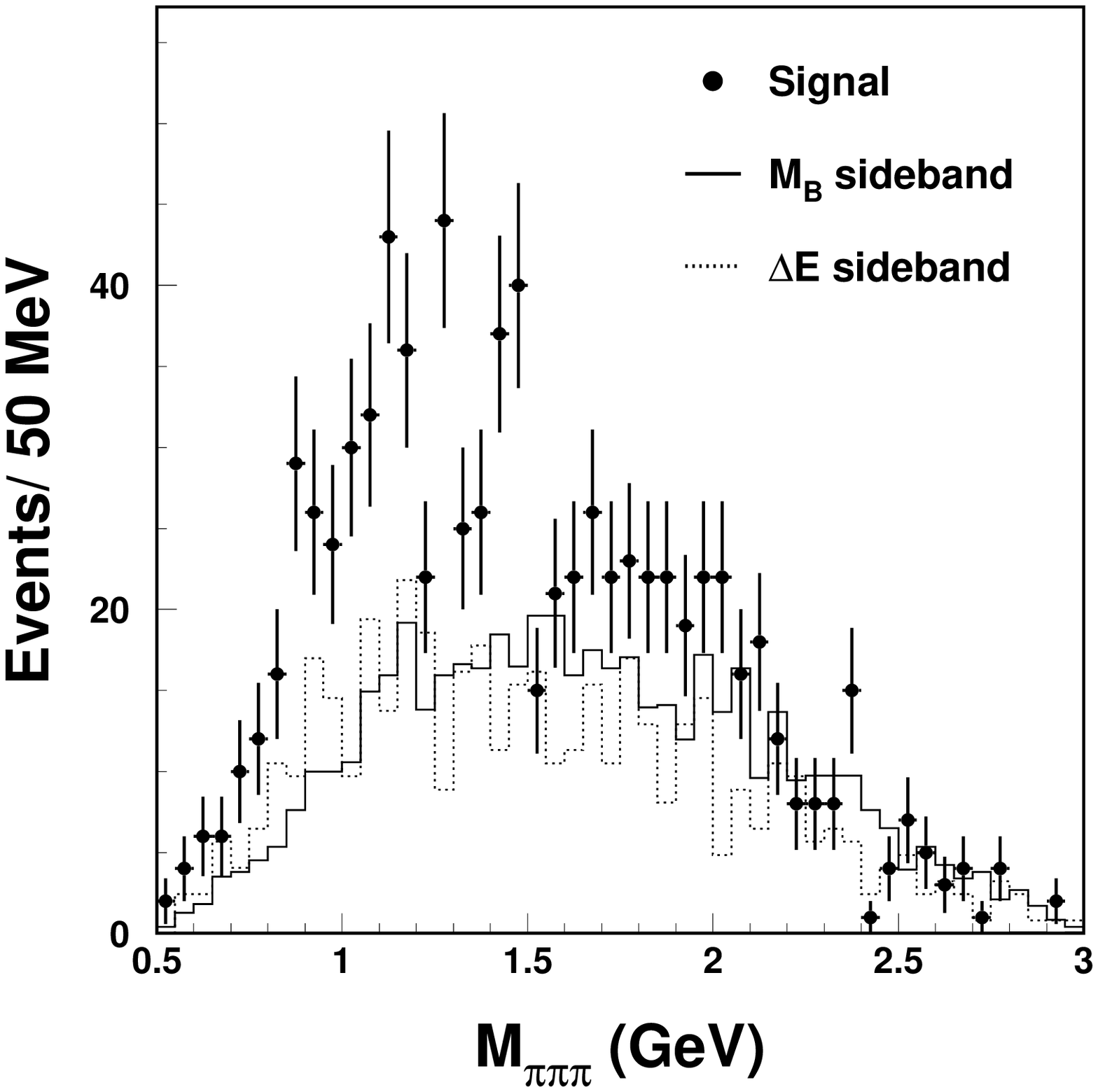,height=4in}}
\vspace{-1.0cm}
\caption{ \label{m3pi_a1_kpi}The invariant mass spectra of 
$\pi^+\pi^-\pi^-$ for the final state
$D^{*+}\pi^+\pi^-\pi^-\pi^o$ for $D^o\to K^-\pi^+$.
The solid histogram is the background
estimate from the $M_B$ lower sideband and the dashed histogram is 
from the
$\Delta E$ sidebands; both are normalized to the fitted number of 
background
events.}
\end{figure}
\section{Search For Other Resonant Substructure in $D^*(4\pi)^-$}
\label{sec:nullsearch}

We have accounted for $\sim$20\% of the $(4\pi)^-$ final state.
We would like to disentangle other resonant substructure.
 Since the background
is large in modes other than $D^o\to K^-\pi^+$ we will only use 
this mode.
One process that comes to mind is that where the virtual $W^-$ 
materializes
as an $a_1^-$, that subsequently decays into $\pi^+\pi^-\pi^-$ and 
we
produce a $D^{**+}$ that decays into $D^{*+}\pi^o$. This process 
should be the
similar to that previously seen in the reaction $B^-\to 
D^{**o}\pi^-$, where
the $D^{**o}$ decayed into a $D^{*+}\pi^-$ \cite{Ddoublestar}.
We search for the presence of an $a_1^-$ by examining
 the $\pi^+\pi^-\pi^-$ mass spectrum in
Fig.~\ref{m3pi_a1_kpi}.

\begin{figure}[H]
\vspace{-0.5cm}
\centerline{\epsfig{figure=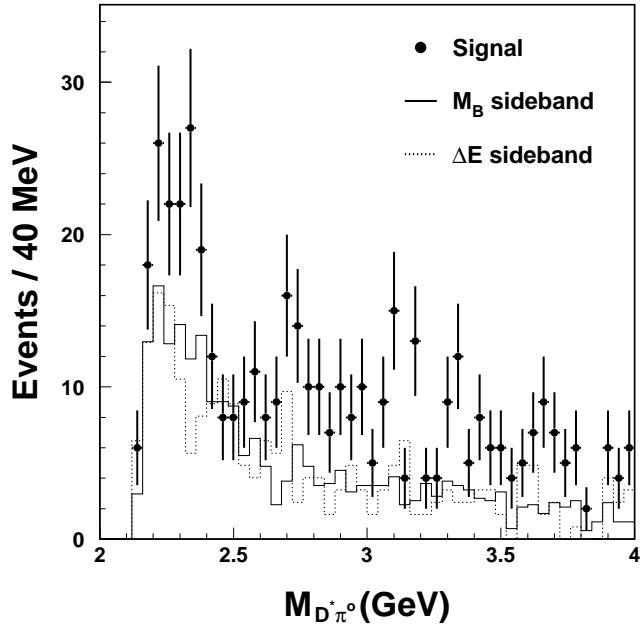,height=4in}}
\vspace{-1.0cm}
\caption{ \label{mdspi0_a1_kpi}The invariant mass spectra of 
$D^{*+}\pi^o$
for $\pi^+\pi^-\pi^-$ masses between 0.6 - 1.6 GeV for the final 
state
$D^{*+}\pi^+\pi^-\pi^-\pi^o$ with $D^o\to K^-\pi^+$.
The solid histogram is the background
estimate from the $M_B$ lower sideband and the dashed histogram is 
from the
$\Delta E$ sidebands; both are normalized to the fitted number of 
background
events.}
\end{figure}

There is an excess of signal events above background in the $a_1^-$ mass
region, that cannot be definitely associated with the $a_1$.
Proceeding by selecting events with $\pi^+\pi^-\pi^-$ masses 
between 0.6
and 1.6 GeV, we show the $D^{*+}\pi^o$ invariant mass spectrum in
Fig.~\ref{mdspi0_a1_kpi}.

\afterpage{\clearpage}
Although there is a suggestion of a low mass enhancement, it is 
not consistent
with $D^{**}$ production that would peak in region of 2.42 - 2.46 
GeV. Perhaps
we are seeing an indication of fragmentation at the $b\to c$ decay 
vertex here.

\begin{figure}[htb]
\centerline{\epsfig{figure=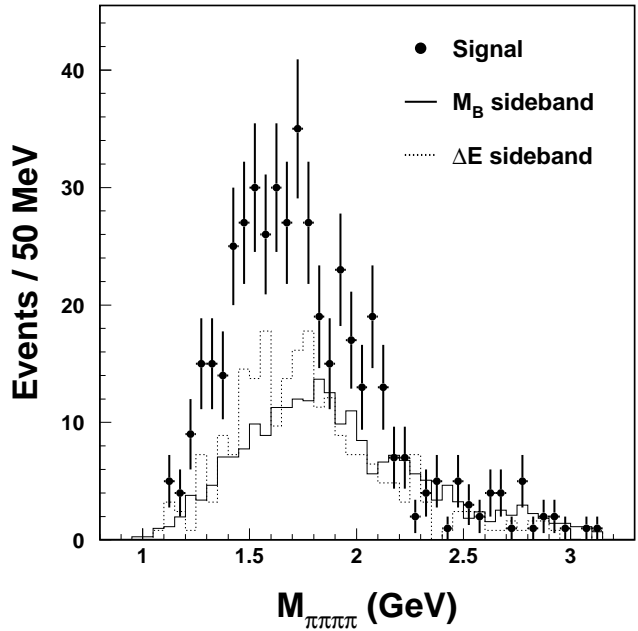,height=4in}}
\vspace{-1.0cm}
\caption{ \label{m4pi_a1_kpi}The invariant mass spectra of
$\pi^+\pi^-\pi^-\pi^o$
for $\pi^+\pi^-\pi^-$ masses between 0.6 - 1.6 GeV for the final 
state
$D^{*+}\pi^+\pi^-\pi^-\pi^o$ with $D^o\to K^-\pi^+$.
The solid histogram is the background
estimate from the $M_B$ lower sideband and the dashed histogram is 
from the
$\Delta E$ sidebands; both are normalized to the fitted number of 
background
events.}
\end{figure}

We also display for completeness the ``$a_1^-\pi^o$" mass 
distribution in
Fig.~\ref{m4pi_a1_kpi}.
There may or may not be a wide structure in the $(4\pi)^-$ mass. 
At this point
we abandon our search for substructure in this decay channel.

\section{Conclusions}\label{sec:conclusions}

We have made the first statistically significant observations of 
six hadronic $B$ decays shown in Table~\ref{table:brs}. 

\begin{table}[hbt]
\begin{center}
\caption{Measured Branching Ratios}
\label{table:brs}
\begin{tabular}{lcr}\hline\hline
Mode & $\cal{B}$ (\%) & \# of events\\\hline
$\overline{B}^o\to D^{*+}\pi^+\pi^-\pi^-\pi^o$ & 
1.72$\pm$0.14$\pm$0.24   &    1230$\pm$70  \\
$\overline{B}^o\to D^{*+}\omega\pi^-$ & 0.29$\pm$0.03$\pm$0.04& 136$\pm$15  
\\
$\overline{B}^o\to D^{+}\omega\pi^-$ & 0.28$\pm$0.05$\pm$0.03     
&  91$\pm$18  \\
${B}^-\to D^{*o}\pi^+\pi^-\pi^-\pi^o$&1.80$\pm$0.24$\pm$0.25       
&    195$\pm$26   \\
${B}^-\to D^{*o}\omega\pi^-$  & 0.45$\pm$0.10$\pm$0.07       &      
26$\pm$6 \\
${B}^-\to D^{o}\omega\pi^-$  &0.41$\pm$0.07$\pm$0.04    & 
88$\pm$14     \\
\hline
\end{tabular}
\end{center}
\end{table}

There is a low-mass resonant substructure in the $\omega\pi^-$ 
mass. A simple Breit-Wigner fit assuming a single resonance and no 
background gives a mass of 1418$\pm$26$\pm$19 MeV with an intrinsic width 
of 382$\pm$41$\pm$32 MeV. 

 The structure at 1418 MeV has a spin-parity 
consistent with $1^-$. It is likely to be the elusive $\rho'$ 
resonance \cite{Clegg}. 
These are by far the most accurate and 
least model dependent measurements of the $\rho'$ parameters.
The $\rho'$ dominates the final state. (Thus the branching ratios 
for the $D^{(*)}\omega\pi^-$ apply also for $D^{(*)}\rho'^-$.)

Heavy quark symmetry predicts equal partial widths for $D^*\rho'$ and $D\rho'$. We measure the relative rates to be
\begin{equation}
{{\Gamma\left(\overline{B}^o\to D^{*+} \rho'^- \right)}\over
{\Gamma\left(\overline{B}^o\to D^{+}\rho'^-\right)}} = 1.04 \pm 
0.21 \pm 0.06 
\end{equation}
\begin{equation}
{{\Gamma\left({B}^-\to D^{*o}\rho'^-\right)}\over
{\Gamma\left({B}^-\to D^{o}\rho'^-\right)}} = 1.10 \pm 0.31 \pm 
0.06 
\end{equation}
\begin{equation}
{{\Gamma\left({B}\to D^{*}\rho'^-\right)}\over
{\Gamma\left({B}\to D\rho'^-\right)}} = 1.06 \pm 0.17 \pm 0.04~~~.
\end{equation}

Thus the prediction of heavy quark symmetry is satisfied within 
our errors. 

Factorization predicts that the fraction of longitudinal  
polarization of the $D^{*+}$ is the same as in the related 
semileptonic decay $B\to D^*\ell^-\bar{\nu}$ at four-momentum 
transfer $q^2$ equal to the mass-squared of the $\rho'$
\begin{equation}
{{\Gamma_L\left({B}\to D^{*+}\rho'^-\right)}\over 
{\Gamma\left({B}\to D^{*+}\rho'^-\right)}} = 
{{\Gamma_L\left({B}\to D^{*}\ell^-\bar{\nu}\right)}\over 
{\Gamma\left({B}\to D^{*}\ell^-
\bar{\nu}\right)}}\left|_{q^2=m^2_{\rho'}}\right.~~.
\end{equation}

Our measurement of the $D^{*+}$ polarization (see Fig.~\ref{cosd}) is 
 (63$\pm$9)\%. 
The model predictions in semileptonic decays for a $q^2$ of 2 
GeV$^2$, are between
66.9 and 72.6\% \cite{slmodels}. Thus this prediction of factorization is 
satisfied. 

We can use factorization to estimate the product of the $\rho'$ 
decay constant $f_{\rho'}$ 
and the branching ratio for $\rho'^-\to\omega\pi^-$. The relevant 
expression is
\begin{equation}
{{\Gamma\left(B\to D^{*+}\rho'^- ,~\rho'^-\to\omega\pi^-\right)}\over 
{{d\Gamma \over dq^2}\left(B\to D^*\ell^-\nu\right)|_{q^2=m^2_{\rho'}}}}
=6\pi^2c_1^2f^2_{\rho'}{\cal{B}}\left(\rho'^-\to\omega\pi^-\right)|V_{ud}|^2~~,
\end{equation}
where $f_{\rho'}$ is the so called $\rho'$ decay constant and $c_1$ is a QCD 
correction factor. We use $c_1$=1.1$\pm$0.1 \cite{scaleerr}.

We use the semileptonic decay rates given in Barish \etal ~\cite{Barish}.
The product 
\begin{equation}
f_{\rho'}^2{\cal{B}}\left(\rho'^-\to\omega\pi^-\right)=0.011\pm 0.003 {\rm
~GeV^2}~~,
\end{equation} 
where the error is the quadrature of the experimental errors on the
experimental branching ratios and $c_1$.

The model of Godfrey and Isgur predicts decay constants widths and partial
widths of mesons comprised of light quarks by using a relativistic treatment
in the context of QCD \cite{GImodel}. They predict both $f_{\rho'}$ and
${\cal{B}}\left(\rho'^-\to\omega\pi^-\right)$; the values are 80 MeV and
39\%, respectively. The branching ratio prediction
is believed to be  more accurate \cite{GInotes}. We use this to extract
\begin{equation}
f_{\rho'}=167   \pm 23 {~\rm MeV}~~.
\end{equation}
The model predicts a lower value for $f_{\rho'}$ observed here, if 
factorization is correct.

We note that all the $B\to D^{(*)}\rho'$ branching ratios that we have 
measured are approximately equal to the $B\to D^{(*)}\rho$ branching rates 
\cite{BigB}
if a model value of ${\cal{B}}\left(\rho'^-\to\omega\pi^-\right)$ = 39\% is
used.

Finally, although the $\overline{B}^o\to D^{*+}(4\pi)^-$ and 
${B}^-\to D^{*o}(4\pi)^-$ branching ratios are nearly equal, the 
$\omega\pi^-$ 
branching ratios are about 1.5 times larger for the charged 
$B$ than the 
neutral $B$, maintaining the trend seen for the $\pi^-$ and 
$\rho^-$ final states. 
Since the $B^-$ lifetime is if anything longer than the $B^o$, 
this trend 
must reverse for some final states. It has not for $D^{(*)}\rho'$.

\section{Acknowledgements}
We thank A. Donnachie, N. Isgur and J. Rosner for useful discussions.
We gratefully acknowledge the effort of the CESR staff in providing us with
excellent luminosity and running conditions.
This work was supported by 
the National Science Foundation,
the U.S. Department of Energy,
the Research Corporation,
the Natural Sciences and Engineering Research Council of Canada, 
the A.P. Sloan Foundation, 
the Swiss National Science Foundation, 
the Texas Advanced Research Program,
and the Alexander von Humboldt Stiftung.


\begin{thebibliography}{99}
\bibitem{Differences}
We do not find significant differences between
the CLEO II and II.V data sets.
\bibitem{spineff}
We assume that $\omega\pi$ is in a $1^-$ final state. The 
difference
between the $1^-$ efficiency and flat assumption is less than 2\%.
\bibitem{systematics} We allowed a systematic error of 5.5\% on 
the number of
events in each angular bin resulting possibly from the fitting 
procedure
to the $M_B$ spectrum. This changed the probability of $\chi^2$ 
for the 
$1^-$ case to 3.9\% from 3.8\%, and negligible changes in the 
other cases.
\bibitem{GInotes}
Private communication from N. Isgur.
The decay constants given in ref.~\cite{GImodel} are dimensionless and must
be multiplied by $m$ to put them in the convention used here.
An additional a factor of $\sqrt{2}$ must be used to
convert from the electromagnetic form-factor to the weak form-factor using
CVC.
\end{thebibliography}
\end{document}